\documentclass[%
 reprint,
 nofootinbib,
 amsfonts,amsmath,amssymb,
 aps,
 prd,
]{revtex4-2}

\usepackage[dvipsnames]{xcolor}

\usepackage[a-2u]{pdfx}
\hypersetup{unicode}
\hypersetup{breaklinks=true}
\hypersetup{
    colorlinks=true,
    urlcolor=blue,
    linkcolor=blue,
    citecolor=blue
}

\everymath=\expandafter{\the\everymath\displaystyle}

\usepackage{amsmath}        
\usepackage{amsfonts}       
\usepackage{amsthm}         
\usepackage{amssymb}

\usepackage{lmodern}
\usepackage{csquotes}
\usepackage[english]{babel}

\usepackage[dvipsnames]{xcolor}

\usepackage{graphicx}

\usepackage{microtype}

\usepackage{dcolumn}
\usepackage{booktabs}

\usepackage{multirow}

\usepackage[caption=false]{subfig}

\usepackage{upgreek}
\usepackage[scr=boondoxo]{mathalpha}
\DeclareMathAlphabet{\mathsfit}{T1}{\sfdefault}{\mddefault}{\sldefault}

\newcommand{\figinput}[1]{\includegraphics{#1.pdf}}%
\newcommand{\colorNP}{blue}%
\newcommand{\colorK}{red}%
\newcommand{\colorC}{olive}%
\definecolor{colorKN}{HTML}{008e95}
\definecolor{fig_blue}{RGB}{36, 106, 174} 
\definecolor{fig_red}{RGB}{182, 31, 46} 

\def\pd{\boldsymbol{\partial}}

\newcommand{\abs}[1]{\lvert{#1}\rvert}

\newcommand{\horeq}{\doteq}

\newcommand{\myoverbar}[1]{\mkern 2.0mu\overline{\mkern-2.0mu#1\mkern-0.7mu}\mkern 0.7mu}
\newcommand{\cconj}[1]{\myoverbar{#1}}

\let\oldepsilon=\epsilon
\def\epsilon{\varepsilon}

\newcommand{\NN}{\mathcal{N}}
\renewcommand{\SS}{\mathcal{S}}

\newcommand{\Kv}    {{\color{\colorK}{u}}}
\newcommand{\Kr}    {{\color{\colorK}{s}}}
\newcommand{\KrMM}  {\left(\Kr + 2\massBH\right)}
\newcommand{\Ktheta}{{\color{\colorK}{\vartheta}}}
\newcommand{\Kphi}  {{\color{\colorK}{\varphi}}}
\newcommand{\Kzeta} {{\color{\colorK}{\varsigma}}}

\newcommand{\KzetaE}{\Kzeta_{\kern 0.08em \text{E}}}

\newcommand{\Salpha}    {{\color{\colorNP}{\alpha}}}
\newcommand{\Sbeta}     {{\color{\colorNP}{\beta}}}
\newcommand{\Smu}       {{\color{\colorNP}{\mu}}}
\newcommand{\Snu}       {{\color{\colorNP}{\nu}}}
\newcommand{\Srho}      {{\color{\colorNP}{\rho}}}
\newcommand{\Ssigma}    {{\color{\colorNP}{\sigma}}}
\newcommand{\Skappa}    {{\color{\colorNP}{\kappa}}}
\newcommand{\Stau}      {{\color{\colorNP}{\tau}}}
\newcommand{\Sgamma}    {{\color{\colorNP}{\gamma}}}
\newcommand{\Slambda}   {{\color{\colorNP}{\lambda}}}
\newcommand{\Sepsilon}  {{\color{\colorNP}{\epsilon}}}
\newcommand{\Spi}       {{\color{\colorNP}{\pi}}}
\newcommand{\Sa}        {{\color{\colorNP}{a}}}

\newcommand{\NPl}       {{\color{\colorNP}{\ell\kern 0.02em}}}
\newcommand{\NPn}       {{\color{\colorNP}{n}}}
\newcommand{\NPm}       {{\color{\colorNP}{m}}}

\newcommand{\NPPsi}     {{\color{\colorNP}{\mathit{\Psi} \kern -0.08em}}}
\newcommand{\NPPhi}     {{\color{\colorNP}{\mathit{\Phi} \kern -0.04em}}}

\newcommand{\NPLambda}  {{\color{\colorNP}{\kern -0.08em \mathit{\Lambda}}}}

\newcommand{\TU}        {{\color{PineGreen}{U}}}
\newcommand{\TX}        {{\color{PineGreen}{X}}}
\newcommand{\TOmega}    {{\color{PineGreen}{\mathit{\Omega}}}}
\newcommand{\Txi}       {{\color{PineGreen}{\xi}}}

\newcommand{\cosmConst} {{\color{\colorC}{{}^\text{c}\kern -0.25em \mathit{\Lambda}}}}
\newcommand{\area}      {{\color{\colorC}{A}}} 
\newcommand{\energy}    {{\color{\colorC}{E}}} 
\newcommand{\massBH}    {{\color{\colorC}{M}}} 

\newcommand{\EuclidR}   {{\color{\colorC}{\mathcal{R}}}} 

\newcommand{\Cpi}       {\kern 0.04em \uppi \kern 0.04em} 
\newcommand{\dss}       {{\d\mathscr{s}^2}} 

\newcommand{\aMomentum} {{\color{\colorC}{L}}} 
\newcommand{\surfkappa} {{\color{\colorC}{\kappa}}}

\newcommand{\CarterK}   {{{\color{\colorC}{K}}}} 
\newcommand{\Cartereta} {{{\color{\colorC}{\eta}}}} 

\newcommand{\KNa}       {{\color{\colorC}{\mathsfit{a}}}}
\newcommand{\KNv}       {{{\color{colorKN}{\mathsf{v}}}}} 
\newcommand{\KNr}       {{{\color{colorKN}{r}}}} 
\newcommand{\KNrp}      {{{\color{\colorC}{r_\text{p}}}}} 
\newcommand{\KNrm}      {{{\color{\colorC}{r_\text{m}}}}}
\newcommand{\KNrmSQ}    {{{\color{\colorC}{r}}_{\text{\color{\colorC}{m}}}^{\hspace{0.2em} 2}}}
\newcommand{\KNtheta}   {{{\color{colorKN}{\theta}}}} 
\newcommand{\KNthetaZ}  {{{\color{colorKN}{\theta_\text{p}}}}} 
\newcommand{\KNphi}     {{{\color{colorKN}{\phi}}}} 
\newcommand{\KNEx}      {{{\color{colorKN}{x}}}}
\newcommand{\KNEz}      {{{\color{colorKN}{z}}}}
\newcommand{\KNR}       {{\color{colorKN}{R}}}
\newcommand{\KNP}       {{\color{colorKN}{P}}}
\newcommand{\KNS}       {{\color{colorKN}{S}}}
\newcommand{\KKr}       {{\color{colorKN}{\mathbb{K}_r}}}
\newcommand{\KKtheta}   {{\color{colorKN}{\mathbb{K}_\theta}}}

\newcommand{\KNrho}     {{\color{colorKN}{\varrho}}}
\newcommand{\KNDelta}   {{\color{colorKN}{\mathit{\Delta}}}}
\newcommand{\KNSigma}   {{\color{colorKN}{\mathit{\Sigma}}}}
\newcommand{\KNSigmaZ}  {{\color{colorKN}{{\mathit{\Sigma}}_0}}}

\newcommand{\KNSr}      {\Kr}
\newcommand{\KNStheta}  {\Ktheta}

\newcommand{\KNSEx}     {{{\color{\colorK}{\hat{x}}}}}
\newcommand{\KNSEz}     {{{\color{\colorK}{\hat{z}}}}}
\newcommand{\CCarterK}  {C^{\CarterK}}
\newcommand{\CKr}       {C^{\Kr}}
\newcommand{\CJ}        {C^{\tilde{\Kphi}}}

\newcommand{\caCarterK} {c^{\CarterK}}
\newcommand{\caKr}      {c^{\Kr}}
\newcommand{\caJ}       {c^{\tilde{\Kphi}}}

\newcommand{\KIN}       {\text{{K}}}
\newcommand{\Hm}        {\circ}
\newcommand{\Hmm}       {\circ\circ}
\newcommand{\NPP}       {\text{{NPP}}}
\newcommand{\PP}        {\text{{PP}}}
\newcommand{\PPm}       {\bullet}
\newcommand{\PPmm}      {\bullet\bullet}
\newcommand{\PPmmm}     {\bullet\bullet\bullet}
\newcommand{\GEO}       {\text{{G}}}

\newcommand{\er}        {\oldepsilon_r}
\newcommand{\etheta}    {}
\newcommand{\ethetaD}   {\oldepsilon_\theta}

\newcommand{\Killing}   {k}

\newcommand{\Jphi}      {{{{\tilde{\Kphi}}}}}
\newcommand{\obs}       {\text{o}}
\newcommand{\source}    {\text{s}}
\newcommand{\minoT}     {\tau_{\text{M}}}
\newcommand{\ru}       {\mathsfit{u}}

\newcommand{\OSexp}{\hat{\mathit{\Theta}}}

\DeclareMathOperator{\sign}{sign}

\DeclareMathOperator{\dn}{dn}
\let\Re\relax
\let\Im\relax
\DeclareMathOperator\Re{Re}
\DeclareMathOperator\Im{Im}

\newcommand{\bigo}{\mathcal{O}}
\def\eu{\mathrm{e}}
\def\ii{\mathrm{i}}

\usepackage[normalem]{ulem}

\usepackage[colorinlistoftodos]{todonotes}
\let\oldtodo\todo
\DeclareRobustCommand{\todo}[1]{
\tikzexternaldisable%
\oldtodo[color=red!70,inline]{#1}%
\tikzexternalenable%
}

\newcommand{\remove}[1]{}
\newcommand{\add}[1]{#1}
\newcommand{\eqc}{}

\def\refjnl#1{#1}
\def\prd{\refjnl{Physical Review D}}                
\def\prl{\refjnl{Physical Review Letters}}          
\def\physrep{\ref@jnl{Physics Reports}}             

\usepackage{orcidlink}

\def\d{\mathrm{d}}

\renewcommand{\SS}{\mathcal{S}}

\begin{document}

\preprint{APS/123-QED}

\title{
    Kerr isolated horizon revisited: Caustic-free congruence and adapted tetrad
}

\author{Ale\v{s} Flandera\,\orcidlink{0000-0002-9288-8910}}
\email{flandera.ales@utf.mff.cuni.cz}
\author{David Kofro\v{n}\,\orcidlink{0000-0002-0278-7009}}
\email{d.kofron@gmail.com}
\author{Tom\'{a}\v{s} Ledvinka\,\orcidlink{0000-0002-6341-2227}}
\email{tomas.ledvinka@mff.cuni.cz}

\affiliation{
    Institute of Theoretical Physics, Faculty of Mathematics and Physics, 
    Charles University, V Hole\v{s}ovi\v{c}k\'ach 2, 180~00 Prague, Czech Republic
}

\date{\today}

\begin{abstract}
    We revisit the near-horizon description of the Kerr space-time in the
    isolated horizon formalism using a non-twisting null geodesic congruence
    and eliminate the coordinate and geodesic pathologies that arise when the
    Carter constant of motion is globally fixed to a single constant.
    Adopting instead a previously proposed choice of the Carter constant which
    depends on the polar angle on the horizon, we obtain an analytic construction
    of the Newman--Penrose tetrad adapted to isolated horizons together with
    horizon-adapted coordinates in which its defining properties are manifest.
    We compute the associated curvature scalars and provide initial data on
    characteristics for the isolated horizon. In addition to an analytical
    solution, derived by leveraging extensive results on Kerr null geodesics,
    we develop two complementary series expansions and outline a practical
    numerical recipe to make the construction readily usable. Relative to
    earlier treatments, our formulation avoids caustic-induced breakdowns and
    incomplete coordinate coverage while yielding 
    a detailed description of the Kerr black hole in the isolated horizon
    approach.
\end{abstract}

\keywords{general relativity, isolated horizons, deformed Schwarzschild black
hole, accretion disk, perturbation}

\maketitle


\section{Introduction}
    \label{sec:introduction}
    The framework of isolated horizons provides a quasi-local description
    of black-hole boundaries that is independent of the space-time's
    asymptotics, making it ideal for analyzing locally equilibrium settings
    embedded within fully time-dependent space-times.
    Within this approach, the geometry of the horizon is encoded in 
    its intrinsic two-metric and its evolution in an equivalence class $[\NPl^{a}]$.

    Crucially, this formalism establishes quasi-local notions of mass, angular
    momentum, and surface gravity based solely on this intrinsic horizon
    geometry, e.g.\
    \cite{Ashtekar-2000c, Ashtekar-2001, Ashtekar-2002}.
    Given the physical significance of the Kerr solution, the specific
    properties of its intrinsic horizon geometry have been thoroughly studied
    to ensure these quasi-local quantities reproduce standard results, e.g.\
    \cite{Lewandowski-2002, Ashtekar-2002}.
    
    In order to describe the space-time in the vicinity of an isolated
    horizon, Krishnan introduced a Newman--Penrose tetrad together with the
    conditions it must satisfy to represent an isolated horizon, \cite{Krishnan-2012}.
    However, even when an explicit space-time metric is available,
    identifying a tetrad that is adapted to the isolated horizon --- 
    that is, one that satisfies these conditions --- has proven difficult.

    Scholtz constructed such a tetrad for the Kerr--Newman family,
    \cite{Scholtz-2017}, by applying transformations to the well-known 
    Kinnersley tetrad so that it is both \emph{non-twisting} and \emph{parallel-propagated} along its generators. 
    This marked the first fully consistent isolated horizon description of a
    rotating, charged black hole. However, the difficulties involved resulted 
    in a tetrad that was given analytically but not explicitly (except on the horizon).

    Recently, \cite{Kofron-2024} revisited the construction for the uncharged
    Kerr space-time. First, the earlier result of \cite{Scholtz-2017} is completed
    by an explicit expression for the adapted tetrad.
    More importantly,
    it was observed that the original choice of the Carter constant and, hence,
    the null generators from \cite{Scholtz-2017}, leads either to caustics or
    coordinates that do not cover the whole space-time (depending on the
    choice of the other constants of motion).
    Moreover, an improved choice of the Carter constant has been proposed in
    \cite{Kofron-2024}, which, unlike the one from \cite{Scholtz-2017}, is not
    constant.
    Following an extensive computation, analytical expressions for both
    the Carter integral of motion and the transversal vector $\NPn^a$ were
    found in \cite{Kofron-2024} in terms of Weierstrass functions. However, 
    the other tetrad vectors remained unavailable, as did all
    the Newman--Penrose scalars.

    Nevertheless, both \cite{Scholtz-2017} and \cite{Kofron-2024} assume prior
    knowledge of the complete space-time structure: the metric and a
    Newman--Penrose tetrad are taken as given from the outset.
    In contrast, \cite{Flandera-2025} demonstrates that the assumption of
    global spherical symmetry alone is sufficient to yield a particular
    isolated horizon. A natural continuation of this line of work would be to
    build on the refined tetrad introduced in \cite{Kofron-2024} and develop an
    analytic slow-rotation expansion of the Kerr isolated horizon. However,
    one can instead exploit the construction of parallel-propagated frames
    based on the hidden symmetries of Kerr space-times, together with recent
    results on Kerr geodesic congruences, to remove any restriction on the
    value of the Kerr rotation parameter $\KNa$.
    We found the treatment of parallel-propagated frames by Kubiz\v{n}\'{a}k et al.,
    \cite{Kubiznak-2009}, and the compendium of null geodesics by Gralla and 
    Lupsasca, \cite{Gralla-2020}, useful.
    Thus, in this paper, we 
    present a construction of the non-twisting null congruence 
    defined by its properties on the horizon 
    that aligns with the geometry of the Kerr black hole space-time
    in the isolated horizon approach.

    Since we build upon two previous papers, we do not repeat the detailed
    introduction to isolated horizons and Newman--Penrose formalism. Instead,
    we only briefly describe the Newman--Penrose quantities as they appear in the
    text, and we refer the reader to 
    \cite{Scholtz-2017, Krishnan-2012, Kofron-2024}
    for further details.

    The paper is structured as follows: In Sec.~\ref{sec:congruences},
    we introduce a \emph{general null geodesic congruence}, establish most of
    the notation used, and describe the choice of the Carter constant of motion
    proposed by \cite{Kofron-2024}.
    Section~\ref{sec:non-twisting_tetrad} constructs the Kerr black hole \emph{tetrad
    adapted to isolated horizons} in Kerr null coordinates. For most of the construction,
    we consider a general Carter constant of motion and employ the particular
    choice only where necessary.
    We discuss the connection between the adapted tetrad and the Kinnersley tetrad and
    derive the initial data on characteristics (the horizon and a transversal hypersurface)
    for the isolated horizon.
    Section~\ref{sec:coordinates} presents \emph{coordinates adapted to the isolated horizon}
    \cite{Krishnan-2012} and the corresponding form of the Newman--Penrose tetrad.
    Moreover,
    the conditions of an axial isolated horizon, as defined by \cite{Ashtekar-2004},
    are discussed. 
    Section~\ref{sec:particular_solutions} focuses on previously implicitly defined
    functions, including the \emph{Carter constant} and \emph{the affine parameter}
    of the geodesics. They are first analyzed analytically, producing complex
    expressions, and then approximated via two different series expansions: 1)
    in the radial coordinate around the horizon and 2) in the rotational parameter $\KNa$.
    We also provide a recipe for a numerical solution.

    To support transparency and reproducibility, the results presented in this work
    are accompanied by a set of Wolfram Mathematica notebooks that document selected
    underlying calculations, intermediate steps, and numerical evaluations.
    These materials are made available to the reader through an openly accessible
    repository hosted on Zenodo, \cite{zenodo-data}.
    The notebooks allow independent verification of the results and provide a
    practical resource for reproducing and extending the analyses reported in this
    paper.

\section{Non-twisting null geodesic congruence}\label{sec:congruences}
    The Kerr metric in null coordinates, that is regular at the horizon,
    takes the form, \cite{Poisson-2004}:%
    \footnote{
        In order to make a clear distinction between different coordinates
        and symbols, we use the following color map of symbols:
        the (Krishnan) coordinates adapted to the isolated horizon
        are highlighted in {\color{\colorK}{red}} color,
        while Kerr null coordinates (and repeating expressions of them) are
        colored {\color{colorKN}{teal}}. Moreover, the Newman--Penrose quantities,
        such as the tetrad vectors, derivatives, and scalars, are marked in
        {\color{\colorNP}{blue}} and metric functions {\color{PineGreen}{green}}.
        Finally, {\color{\colorC}{olive}} is used for
        physical constants.

    }
    \begin{align}
        \begin{split}
            \dss &=
            \left(1 - \frac{2\massBH \KNr}{\KNSigma}\right) \d\KNv^2
            - 2\,\d\KNv\,\d\KNr \\
            &\qquad{}+ \frac{4\KNa \massBH\KNr \sin^2\KNtheta}{\KNSigma} \,\d\KNv\,\d\KNphi 
            + 2 \KNa \sin^2\KNtheta \,\d\KNr\,\d\KNphi \\
            &\qquad{}- \KNSigma \,\d\KNtheta^2 
            + \frac{\sin^2\KNtheta}{\KNSigma}\! \left(\!\KNDelta \KNa^2 \sin^2 \KNtheta - \KNSigmaZ^2\right) \d\KNphi^2 \,,
        \end{split}
        \label{eq:Kerr_null_coordinates}
    \end{align}
    where
    \begin{align}
        \KNDelta &= \KNr^2 - 2\massBH\KNr + \KNa^2 \,, \\
        \KNrho &= \KNr + \ii \KNa \cos \KNtheta \,, \\
        \KNSigma &= \KNr^2 + \KNa^2 \cos^2\KNtheta \,, \\
        \KNSigmaZ &= \KNr^2 + \KNa^2 \,.
    \end{align}

In these coordinates, the general null
geodesic congruence $\NPn^a_\GEO$ implied by the Carter equations
\cite{Carter-1968} has components
\begingroup
\allowdisplaybreaks
\begin{subequations}
\begin{align}
    \NPn^\KNv_\GEO      &= - \frac
    {\KNa^2 \energy \sin^2 \KNtheta + \KNa \aMomentum + \frac{\KNSigmaZ}{\KNDelta}
    \left(\sqrt{\KNR} - \KNP\right)}{\KNSigma} \,, \\
    \NPn^\KNr_\GEO      &=  \er\, \frac{\sqrt{\KNR}}{\KNSigma}  \,, \label{eq:vector_n_general_r}\\
    \NPn^\KNtheta_\GEO  &=  \ethetaD\, \frac{\sqrt{\KNS}}{\KNSigma}  \,, \\
    \NPn^\KNphi_\GEO    &= - \frac{\KNa \energy + \aMomentum \sin^{-2}\KNtheta + \frac{\KNa}{\KNDelta}
    \left(\sqrt{\KNR} - \KNP\right)}{\KNSigma} \,,
\end{align}
\label{eq:vector_n_general}
\end{subequations}
\endgroup
where
\begin{align}
    \KNP &= \KNa \aMomentum + \energy \KNSigmaZ \,, \\
    \KNR &= \KNP^2 - \CarterK \KNDelta \,, \label{eq:radialPolynomialR}\\
    \KNS &= \CarterK - \left(\aMomentum + \KNa \energy\right)^2
    + \left(\KNa^2 \energy^2 - \aMomentum^2 \sin^{-2}\KNtheta\right)
    \cos^2\KNtheta \,,
\end{align}
and $\energy$ is energy, $\aMomentum$ is angular momentum, and
$\CarterK$ is the Carter constant of motion. We consider~\eqref{eq:vector_n_general}
as a space-time vector field rather than one restricted along a particular
null geodesic, so these constants of motion must satisfy
\begin{align}
    \NPn_\GEO^a \nabla_a \energy &= 0 \,, &
    \NPn_\GEO^a \nabla_a \aMomentum &= 0 \,, &
    \NPn_\GEO^a \nabla_a \CarterK &= 0 \,.
\end{align}
Since we want $\NPn^a$ to be inward pointing,
as per \cite{Krishnan-2012}, we need to choose the radial square root sign
as $\er = -1$.

As previously indicated, we consider the \emph{non-twisting}
null geodesic congruence proposed 
by~\cite{Kofron-2024}, characterized by 
\begin{equation}
    \NPn_a^\GEO\, \d x^a = \d \Kv \,,
\end{equation}
where the role of $\Kv$ as a null coordinate will be explained in Sec.%
~\ref{sec:coordinates}. As argued in~\cite{Scholtz-2017}, to obtain a
non-twisting congruence $\NPn^a_\GEO$ that exists everywhere, we have to put
$\aMomentum=0$. Following the suggestion of \cite{Kofron-2024},
we also single out a particular choice of the Carter constant
at the horizon that not only corrects the singular behavior at the axis 
of the congruence chosen in \cite{Scholtz-2017}, but also leads to vanishing
$\NPn^\KNtheta_\GEO$ at the horizon.%
\footnote{Note that, for the Kinnersley tetrad, this component is vanishing everywhere.}
The geodesic starting at the horizon with $\KNr\horeq\KNrp$ and
$\KNtheta\horeq\KNthetaZ$
(by $\horeq$ we denote equality at the horizon and all functions are evaluated
on the horizon, e.g.\ $\KNSigmaZ \horeq \KNrp\!^2 + \KNa^2$)
has
\begin{align}
    \CarterK &= \KNa^2 \sin^2 \KNthetaZ \,.
    \label{eq:K_is_a2_sin2_theta}
\end{align}
Thus, all geodesics in the congruence will have the same $\energy$ and
$\aMomentum$, but each one retains the value of $\CarterK$ it acquires when it
starts at the horizon at $\KNtheta\horeq\KNthetaZ$. Moreover, $\KNthetaZ$
will become the coordinate $\Ktheta$ adapted to the isolated horizon
in Sec.~\ref{sec:coordinates}.

Off the horizon, $\CarterK$ is determined by the requirement of geodesic
congruence
\begin{align}
    \NPn_\GEO^a \nabla_a \CarterK = 0 \,.
    \label{eq:Carter_transport}
\end{align}

With the choice $\energy = 1$ and $\aMomentum = 0$, we have
\begin{align}
    \sqrt{\KNR} \,&\overset{\energy = 1,\, \aMomentum = 0}{=}\, 
        \sqrt{\KNSigmaZ^2 - \CarterK \KNDelta} \,, \\
    \etheta\,\sqrt{\KNS} \,&\overset{\energy = 1,\, \aMomentum = 0}{=}\, 
        \etheta\,\sqrt{\CarterK - \KNa^2 \sin^2 \KNtheta} \,,
\end{align}
and condition~\eqref{eq:Carter_transport} translates to:
\begin{align}
    - \sqrt{\KNSigmaZ^2 - \CarterK \KNDelta}\, \CarterK_{,\KNr}
    + \ethetaD \sqrt{\CarterK - \KNa^2 \sin^2 \KNtheta}\, \CarterK_{,\KNtheta}
    = 0 \,.
    \label{eq:Carter_geodesic_cond}
\end{align}
Obviously, Eq.~\eqref{eq:Carter_geodesic_cond} states that $\CarterK$ is
constant along any single null ray, but it varies over the manifold.
Coordinates $\KNr$, $\KNtheta$ of such a ray starting at the horizon
satisfy the implicit equation
\begin{equation}
    \int\limits_{\KNthetaZ}^\KNtheta
    \frac{\d \KNtheta'}{\ethetaD\sqrt{\CarterK - \KNa^2 \sin^2 \KNtheta'}}
    + \int\limits_{\KNrp}^\KNr  
    \frac{\d \KNr'}{\sqrt{\KNSigmaZ'^2 - \CarterK \KNDelta'}} 
    = 0 \,.
    \label{eq:Carter_implicit1}
\end{equation}
When an explicit dependence of the Carter constant on the horizon coordinate
$\KNthetaZ$, such as $\CarterK=\KNa^2 \sin^2 \KNthetaZ$, is substituted into this equation,
it becomes an implicit equation for $\KNthetaZ(\KNr,\KNtheta)$ and, therefore, also
for $\CarterK(\KNr,\KNtheta)$. It is straightforward to see that we must set
$\ethetaD=\sign(\cos \KNtheta)\sign(\KNr-\KNrp)$ so that the integrals have
opposite signs.
Moreover, the Carter constant must satisfy
\begin{align}
    \CarterK(\KNr, 0) &= \CarterK(\KNr, \Cpi) = 0 \,, &
    \CarterK(\KNr, \tfrac{\Cpi}{2}) &= \KNa^2 \,,
\end{align}
we will discuss the behavior at the axis and equator later
in Sec.~\ref{sec:particular_solutions}. 
We will meet the implicit Eq.~\eqref{eq:Carter_implicit1} expressed in 
terms of elliptic integrals in Sec.~\ref{sec:analytical_solution}.

Because \eqref{eq:Carter_implicit1} involves elliptic integrals,
we cannot write down a closed-form solution for $\CarterK(\KNr,\KNtheta)$.
Nevertheless, note that implicit equations are nothing exceptional;
even simple Newtonian orbital motion requires solving the so-called Kepler
(transcendental) equation either numerically or via series expansions.
We will discuss both approaches in Sec.~\ref{sec:particular_solutions}.

To simplify the following expressions, we define
\begin{align}
    \KKr &\equiv 
    \sqrt{\KNSigmaZ^2 - \CarterK \KNDelta} 
    \,, \\
    \etheta\KKtheta &\equiv 
    \ethetaD\,\sqrt{\CarterK - \KNa^2 \sin^2 \KNtheta}
    \,.
\end{align}
Note that, thanks to the inclusion of $\ethetaD$
in $\KKtheta$, the function has continuous spatial derivatives at $\KNr = \KNrp$
and $\KNtheta = \Cpi/2$.

Apart from the vanishing twist, the expansion of the congruence of $\NPn^a$ is
also worth mentioning. It is given by:
\begin{align}
    \OSexp_\NPn &= \frac{1}{\KNSigma}\Big( \etheta\KKtheta\cot\KNtheta 
    +\etheta \KKtheta_{,\KNtheta} - \KKr_{,\KNr}\Big) 
     \,.
    \label{eq:expansion_of_n}
\end{align}
Note that in the Newman--Penrose formalism, we have
$\OSexp_\NPn \equiv 2\Re\Smu$.
The expansion must be 
finite.
The term $\KKr$ is real and finite for $\KNr\in(0,\infty)$, and so is its derivative.
On the other hand, the regularity of the term $\KKtheta \cot\KNtheta$ on the axis
requires $\KKtheta$ to be $\bigo(\sin\KNtheta)$ (as $\KNtheta \to 0$ and
$\KNtheta \to \Cpi$).
Oftentimes, it is also required that the
expansion is negative, e.g.\ for smooth transitions from a dynamical horizon 
(with this property) to an isolated one (which does not require the property).
This imposes further limitations on the choice of $\CarterK$. See
\cite{Ashtekar-2004LR} for discussion.
Notice that our choice of $\CarterK$ does satisfy all these conditions.

\begin{figure*}[p]
    \subfloat[Kinnersley $\boldsymbol{\NPn}_\KIN$]{
    \begin{tabular}{@{}c@{}c@{}c@{}c@{}}
        \includegraphics[width=.25\textwidth,keepaspectratio]{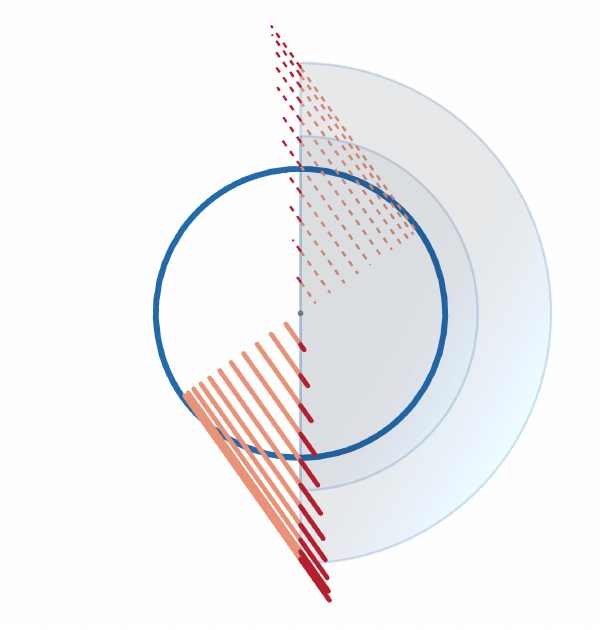} &
        \includegraphics[width=.25\textwidth,keepaspectratio]{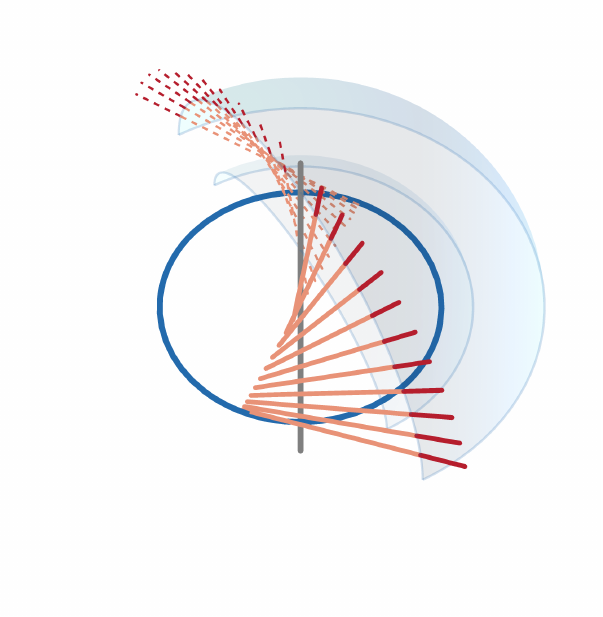} &
        \includegraphics[width=.25\textwidth,keepaspectratio]{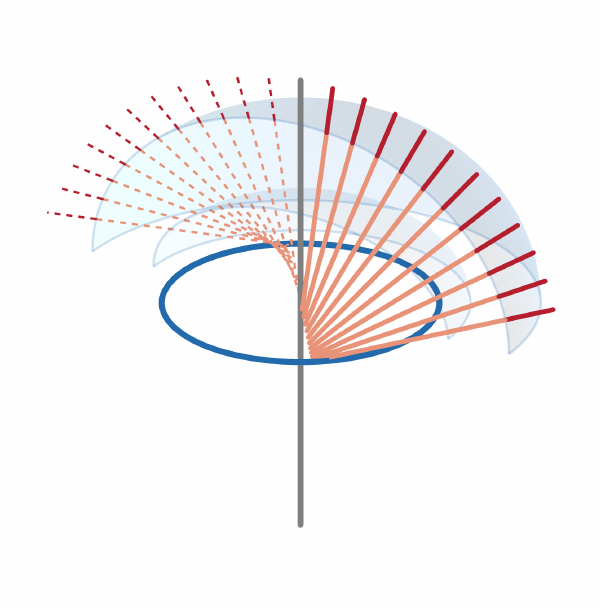} &
        \includegraphics[width=.25\textwidth,keepaspectratio]{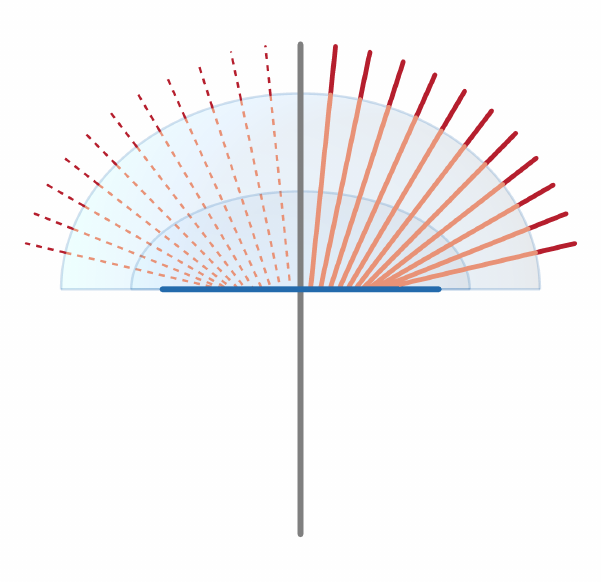} \\
    \end{tabular}
    }\\
    \subfloat[Non-twisting $\boldsymbol{\NPn}_\GEO$ with $\CarterK=\KNa^2$]{
    \begin{tabular}{@{}c@{}c@{}c@{}c@{}}
        \includegraphics[width=.25\textwidth,keepaspectratio]{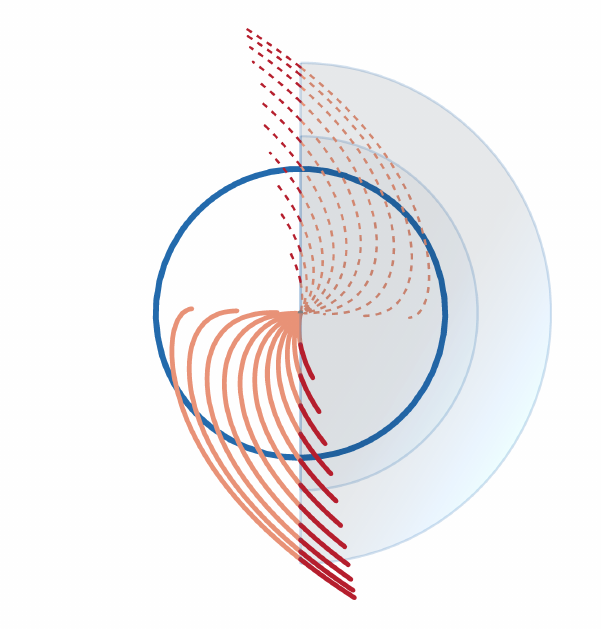} &
        \includegraphics[width=.25\textwidth,keepaspectratio]{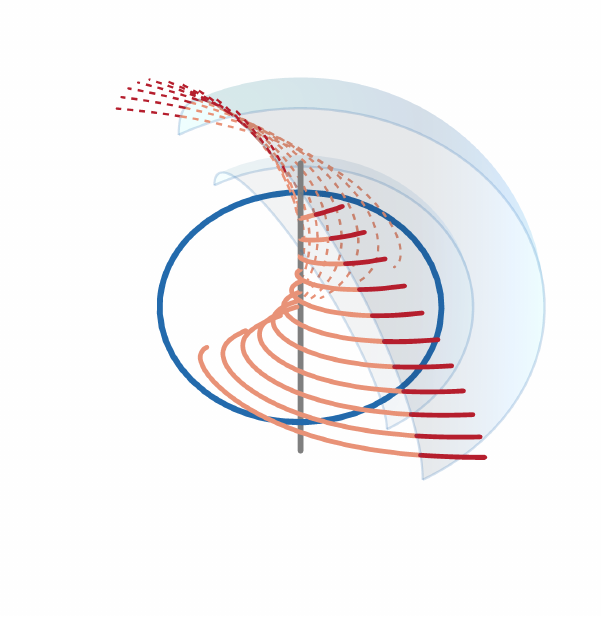} &
        \includegraphics[width=.25\textwidth,keepaspectratio]{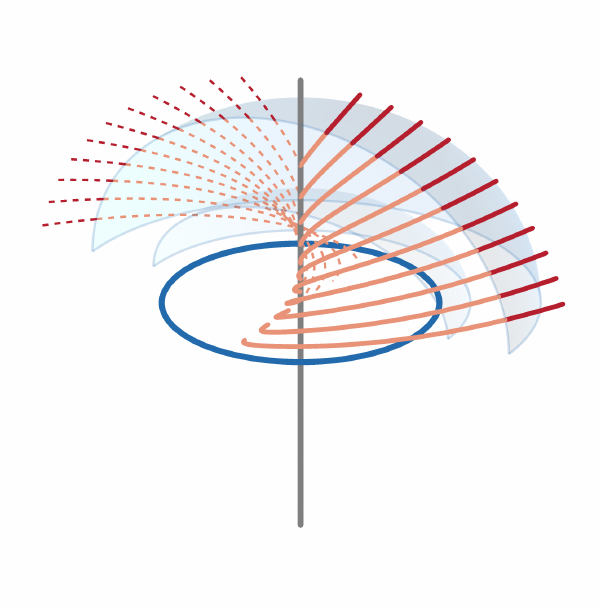} &
        \includegraphics[width=.25\textwidth,keepaspectratio]{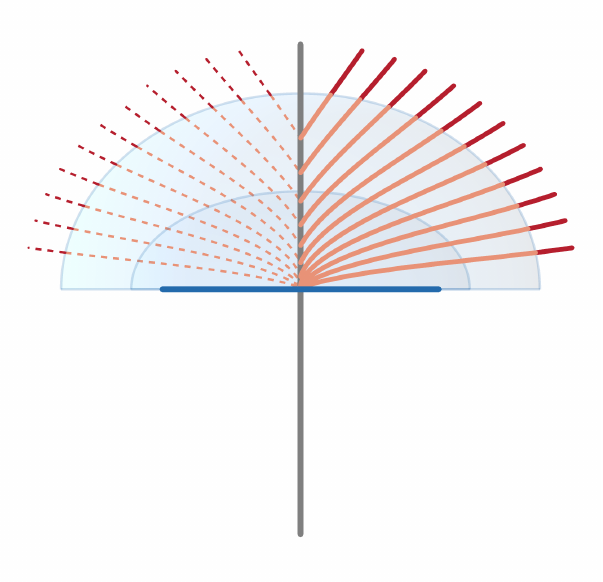} \\
    \end{tabular}
    }\\
    \subfloat[Non-twisting $\boldsymbol{\NPn}_\GEO$ with $\CarterK\horeq\KNa^2\sin^2\KNthetaZ$]{
    \begin{tabular}{@{}c@{}c@{}c@{}c@{}}
        \includegraphics[width=.25\textwidth,keepaspectratio]{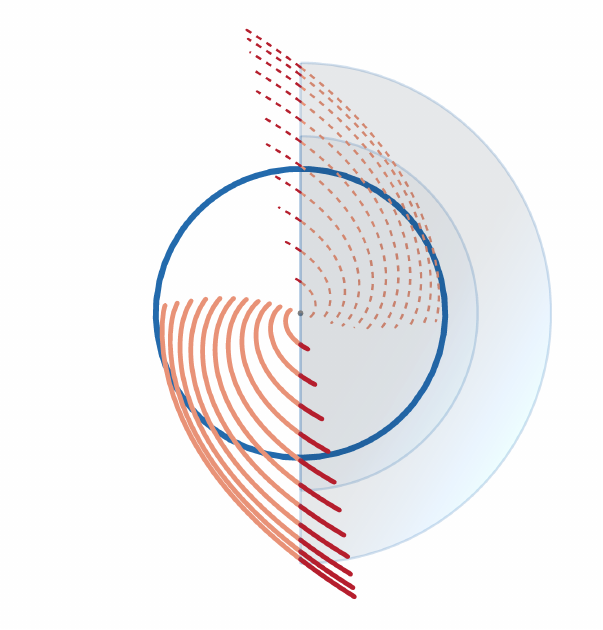} &
        \includegraphics[width=.25\textwidth,keepaspectratio]{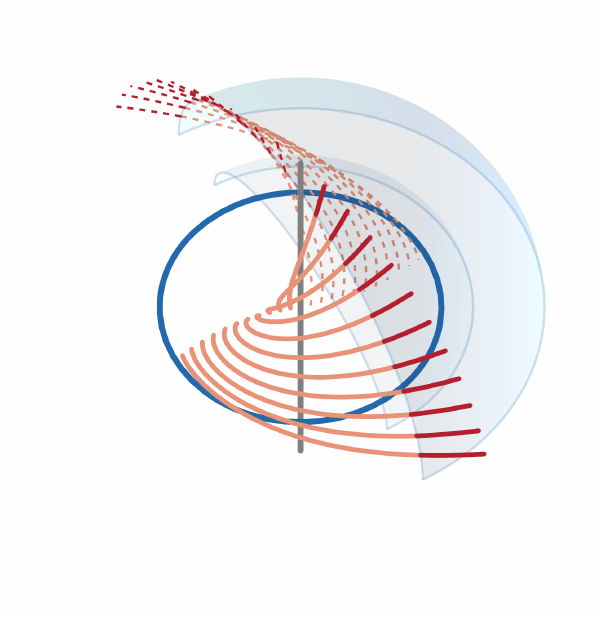} &
        \includegraphics[width=.25\textwidth,keepaspectratio]{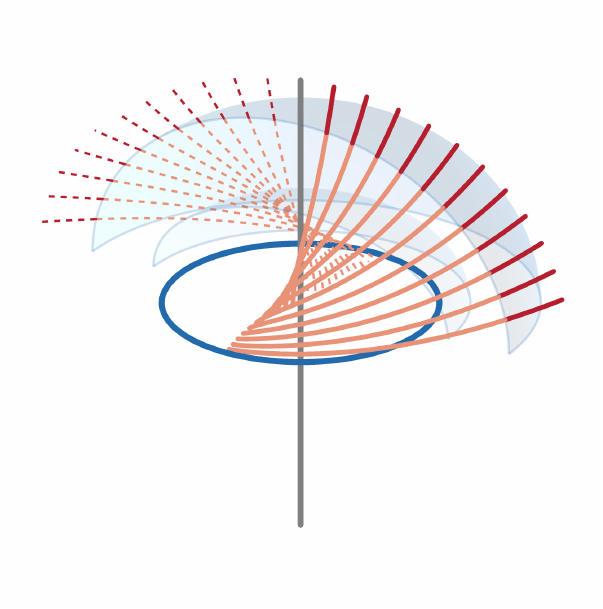} &
        \includegraphics[width=.25\textwidth,keepaspectratio]{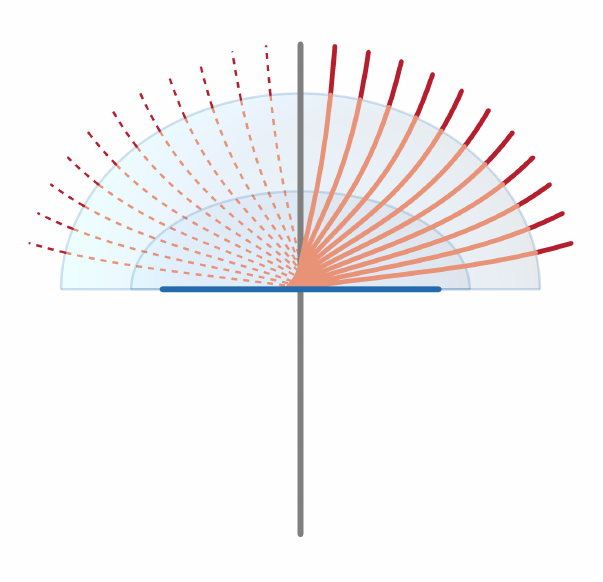} 
    \end{tabular}
    }
    \caption{
        Transversal null geodesics $\NPn^a$ of a Kerr black hole,
        with parameters $\KNrp=2$ and $\KNrm=1$,
        are shown (using a fictitious flat-space projection)
        in Kerr--Schild Cartesian coordinates
        for the Kinnersley tetrad and two non-twisting congruences.
        The grey line marks the rotation axis, the blue circle indicates the ring
        singularity, and the light blue surfaces represent illustrative portions of the inner
        and outer horizons. The geodesics are plotted in red, with lighter 
        shading beneath the outer horizon and darker shading above it.
        Dashed geodesics represent nothing more then geodesics for different
        azimuthal angle.
        In the first row \textbf{(a)}, 
        the twisting geodesic congruence (unsuitable for isolated horizons)
        is shown.
        In the second row \textbf{(b)}, 
        the Carter constant is set to $\CarterK=\KNa^2$,
        producing caustics along the axis. 
        In the third row \textbf{(c)}, 
        the improved
        choice $\CarterK\horeq\KNa^2\sin^2\KNthetaZ$ removes these axis
        caustics and yields geodesics that pass through the disc to the other
        sheet, apart from the obstruction at the ring singularity.
        For each row, the leftmost image gives a view along the axis, the
        rightmost image shows a perpendicular side view, and the central panels
        provide intermediate viewpoints along the camera path connecting the two.
        Supplementary videos providing three-dimensional perspectives of the
        figures are available through an openly accessible
        repository hosted on Zenodo, \cite{zenodo-data}.
    }
    \label{fig:geodesics}
\end{figure*}

To conclude our discussion of the null geodesic congruence, we illustrate
the behavior of the transversal geodesic congruences in Fig.~\ref{fig:geodesics},
which is plotted in the Kerr--Schild Cartesian coordinates;
a reference for these coordinates and the corresponding transformation relations 
can be found in~\cite[Sec.~5.3.7]{Poisson-2004}. 
It can be seen that the improved choice \eqref{eq:K_is_a2_sin2_theta} ensures that
there are no caustics at the axis, while for the simpler choice $\CarterK = \KNa^2$
there are caustics at the axis of rotation.

\section{The non-twisting tetrad}\label{sec:non-twisting_tetrad}
The quantities that describe an isolated horizon are related to a certain
Newman--Penrose tetrad that has been described in detail in \cite{Krishnan-2012}.
Apart from satisfying the usual Newman--Penrose contraction
\begin{align}
    \NPl^a \NPn_a &= 1 \,, &
    \NPm^a \cconj{\NPm}_a &= - 1 \,,
\end{align}
(with all other $=0$), the tetrad must have multiple other properties, see
any of \cite{Krishnan-2012, Scholtz-2017, Kofron-2024} for a review.
Unsurprisingly for a (quasi-)local description of a horizon, most of them are
prescribed at the horizon itself. For this reason, we start by finding the
tetrad on the horizon and extend it off the horizon later.

Although we will use the already thoroughly discussed vector $\NPn^a_\GEO$ and
search for the vectors completing it to the Newman--Penrose frame by employing
the geometrical properties of the final tetrad,
it is practical to also have an existing Newman--Penrose frame as a reference.
In many situations, it is convenient to use the Kinnersley tetrad,%
\footnote{Denoted by (upper or lower) index $\KIN$.}
the Newman--Penrose tetrad adapted to the principal null directions,
\cite{Kinnersley-1969}, which, in the null coordinates, reads
\clearpage 
\begin{subequations}
\begin{align}
    \boldsymbol{\NPl_\KIN} &= \pd_\KNv + \frac{\KNDelta}{2\KNSigmaZ}
    \,\pd_\KNr + \frac{\KNa}{\KNSigmaZ}\, \pd_\KNphi \,, \\
    \boldsymbol{\NPn_\KIN} &= - \frac{\KNSigmaZ}{\KNSigma}\, \pd_\KNr \,, \\
    \boldsymbol{\NPm_\KIN} &= \frac{1}{\sqrt{2} \KNrho} \left( \ii \KNa \sin\KNtheta \,\pd_\KNv
        + \pd_\KNtheta + \frac{\ii}{\sin\KNtheta}\, \pd_\KNphi \right) .
\end{align}
\label{eq:Kinnersley_tetrad}
\end{subequations}
This tetrad has a non-zero twist in the direction of $\NPn^a$ and
is not suitable for describing isolated horizons by itself.  
We will show how this can be remedied by means of Lorentz transformations,
see Appendix~\ref{app:np:lorentz_transformation}, which keep the scalar
products intact while changing other properties of the frame.

\subsection{The horizon}
Let us first discuss a simpler situation. We consider only the restriction of
the tetrad fields on the horizon, where
we can align the vector $\NPn_\KIN^a$ of the Kinnersley tetrad%
~\eqref{eq:Kinnersley_tetrad}
with the non-twisting null geodesic congruence on the horizon
using the following Lorentz transformations:
\begin{enumerate}
    \item a boost with parameter $A_{\KIN\to\Hm}$
        (that changes the Kinnersley tetrad into an intermediate tetrad denoted by the symbol $\Hm$),
    \item a rotation about the new $\NPl^a_\Hm$ with parameter $c_{\Hm\to\Hmm}$
        (so we end up with a tetrad denoted by $\Hmm$).
\end{enumerate}
The parameters are
\begin{align}
    A^2_{\KIN\to\Hm} &= 1 \,, \label{eq:boost_hor}\\
    c_{\Hm\to\Hmm} &= \frac{\ii \KNa \sin\KNtheta + \etheta\KKtheta}{\sqrt{2}\left(\KNrp - \ii \KNa \cos\KNtheta\right)} \,. \label{eq:rot_l_hor}
\end{align}
Notice that the rotation about $\NPl^a$ depends on the choice of the Carter
constant of motion $\CarterK$ (through $\KKtheta$). All the tetrads stemming
from the different choices of this constant are connected by these rotations.

On the horizon, the $\Hmm$-tetrad is identical to the one
adapted to isolated horizons (in the Kerr null coordinates).
Later, we will see tetrads $\PPm$ and $\PPmm$, corresponding to $\Hm$ and $\Hmm$,
which will be set also outside (and inside) the horizon, and we will also express the
tetrad in the Krishnan coordinates adapted to isolated horizons.

Hence, we are looking for a tetrad that, on the horizon, has the form
\begin{subequations}
\begin{align}
    \boldsymbol{\NPl} &\horeq \pd_\KNv + \frac{\KNa}{\KNSigmaZ}\, \pd_\KNphi \,, 
    \label{eq:l_on_H}
    \\[1ex]
    \begin{split}
    \boldsymbol{\NPn} &\horeq \frac{\CarterK - 2 \KNa^2 \sin^2\KNtheta}{2 \KNSigma}\, \pd_\KNv 
    - \frac{\KNSigmaZ}{\KNSigma} \, \pd_\KNr 
    + \frac{\etheta\KKtheta}{\KNSigma} \, \pd_\KNtheta \\
                      &\qquad{}+ \frac{\KNa \left(\CarterK - 2 \KNSigmaZ\right)}{2 \KNSigma \KNSigmaZ} \, \pd_\KNphi \,, \label{eq:nHorizon}
    \end{split}
    \\[1ex]
    \boldsymbol{\NPm} &\horeq \frac{\etheta\KKtheta}{\sqrt{2}\, \KNrho} \, \pd_\KNv
    + \frac{1}{\sqrt{2}\, \KNrho} \, \pd_\KNtheta
    + \frac{\ii \KNSigma \csc\KNtheta + \KNa\etheta\KKtheta}{\sqrt{2}\, \KNrho \KNSigmaZ} \, \pd_\KNphi \,.
\end{align}
\label{eq:tetrad_horizon}
\end{subequations}

Thus, the choice $\CarterK = \KNa^2 \sin^2\KNthetaZ$ proposed
by \cite{Kofron-2024} not only
ensures that $\NPn^\KNtheta \horeq 0$, but also $\NPm^\KNv \horeq 0$.

\subsection{Leaving the horizon}
One of the key characteristics of the isolated horizon describing
the Krishnan tetrad is that the congruence $\NPn^a$ is accompanied by additional
vector fields $\NPl^a$ and $\NPm^a$ that are parallel transported along
$\NPn^a$. We have already employed one miraculous feature of the Kerr geometry that
provided the additional integral of motion~$\CarterK$. However, there are deeper
reasons behind its existence, usually referred to as hidden symmetries of the
Kerr metric \cite{Kubiznak-2009}. In particular, there exists the so-called
Killing--Yano anti-symmetric tensor $f_{ab}$, developed in~\cite{Yano-1952},
that not only provides the Carter constant
$\CarterK=f_{ab} f^b{}_c\, \NPn^c \NPn^b$,
but also allows for the construction of additional parallel transported vector fields.

Using the Kinnersley tetrad~\eqref{eq:Kinnersley_tetrad}, the tensor
is given as, \cite{Kalnins-1989},
\begin{equation}
    f_{ab} = 
    - 2\ii \KNr\, \NPm^\KIN_{[a}\cconj{\NPm}^\KIN_{b]} 
    + 2 \KNa \cos\KNtheta\, \NPl^\KIN_{[a}\NPn^\KIN_{b]} 
    \,,
\end{equation}
where the brackets denote the anti-symmetrization.
Its dual $h^{ab}=\tfrac{1}{2}\,\epsilon^{abcd}f_{cd}$ is similarly
\begin{equation}
    h_{ab} = 
    2\KNr\, \NPl^\KIN_{[a}\NPn^\KIN_{b]} 
    + 2 \ii \KNa \cos\KNtheta\, \NPm^\KIN_{[a}\cconj{\NPm}^\KIN_{b]} 
    \,.
\end{equation}
Provided that $h_{ab}$ is the principal conformal Killing--Yano tensor,
the primary Killing vector $\Killing^a$ is given by
\begin{equation}
    \Killing^a = \frac{1}{3} \nabla_b h^{ba} \,.
\end{equation}

When searching for the parallel-propagated vectors, we could, in principle,
use the same construction as in \cite[Sec.~IV.]{Kofron-2024}.
Namely, to produce the following vectors%
\footnote{
    While the tetrad is NP --- Newman--Penrose, alas,
    it is NPP --- Not Parallel Propagated.
}
\begin{subequations}
\begin{align}
    \NPm^a_\NPP &= \frac{1}{\sqrt{2\CarterK}} \left(
    f^a{}_b\, \NPn^b + \ii\, h^a{}_b\, \NPn^b \right) , \\
    \NPl^a_\NPP &= \frac{1}{\CarterK} \Big(
    f^a{}_b\, f^b{}_c\, \NPn^c + \frac{1}{4}\, f_{bc}\, f^{bc}\, \NPn^a \Big)\, , 
\end{align}%
\label{eq:Kofron-2024_tetrad}%
\end{subequations}%
which are not parallel propagated,
and then perform such a rotation about $\NPn^a$ to acquire parallel 
propagation. Nevertheless, finding the parameter of the rotation 
for $\CarterK$, which is not constant (in the space-time), involves solving
a complicated differential equation.

For this reason, we instead follow the procedure of~\cite{Kubiznak-2009},
which is based on earlier works
\cite{Marck-1983S, Marck-1983P, Kamran-1986} and generalizes them,
and which directly leads to the following parallel-propagated vector fields:
\begin{subequations}
\begin{align}
    \NPm^a_\PP &= - \frac{1}{\sqrt{2\CarterK}}
        \Big[\,\ii\, \big(h^a{}_b + \Kr\, g^a{}_b\big) + f^a{}_b\Big] \, \NPn^b \,,\\
    \NPl^a_\PP &= \frac{1}{2\CarterK}
        \Big[\big(h^a{}_b + \Kr\, g^a{}_b\big) \big(h^b{}_c + \Kr\, g^b{}_c\big)
        + f^a{}_b f^b{}_c\Big]\, \NPn^c \,,
\end{align}%
\label{eq:KY_vectors}%
\end{subequations}%
where the term in parentheses can be understood as the parallel propagator operator.
The scalar field $\Kr$ is required to satisfy, see~\cite{Kubiznak-2009},%
\footnote{We deliberately choose the opposite sign on the right-hand side,
    in comparison to~\cite{Kubiznak-2009}; the rationale for this will become
clear below.}
\begin{equation}%
    \NPn^a \nabla_a \Kr = - \NPn_a\, \Killing^a \,,
    \label{eq:parallel_condition}
\end{equation}%
so that
\begin{align}%
    \NPn^c \nabla_c\! \left(h^a{}_b\, \NPn^b + \Kr\, \NPn^a\right) = 0 \,.
    \label{eq:parallel_propagator}
\end{align}%

In comparison to \cite{Kubiznak-2009}, we have chosen the opposite sign on the
right-hand side of~\eqref{eq:parallel_condition} (and used the anti-symmetry
of the Killing--Yano tensor to factor out the minus in%
~\eqref{eq:parallel_propagator}). The reason is as follows.
For the Kerr space-time, $\Killing^a = \pd_\KNv$, and condition%
~\eqref{eq:parallel_condition} becomes
\begin{equation}
    \NPn^a \nabla_a \Kr = - \energy \overset{\energy = 1}{=} - 1 \,.
    \label{eq:affine_parameter}
\end{equation}
Hence, $\Kr$ is (minus) the affine parameter of the geodesic along $\NPn^a$.
The minus sign has been introduced because, while the vector $\NPn^a$ is inward
pointing, it is desirable for the parameter $\Kr$ to grow outward as a radial
coordinate. Thanks to this, we can identify $\Kr$ with the Krishnan radial
coordinate; see its definition in~\cite{Krishnan-2012}.

It is worth comparing vectors~\eqref{eq:KY_vectors}
to~\eqref{eq:Kofron-2024_tetrad}. Using the following identity
\begin{align}
    f_{ab}f^b{}_c = h_{ab} h^b{}_c + \frac{1}{2} h^{de} h_{de} \,,
\end{align}
we can see that~\eqref{eq:KY_vectors} is corrected by terms depending
on the affine parameter $\Kr$. Moreover, we include a minus sign in the vector
$\NPm^a$ so that its sign convention matches that of the Kinnersley tetrad
(and is consistent with the~\eqref{eq:tetrad_horizon} on the horizon).

Before evaluating vectors~\eqref{eq:KY_vectors}, let us rewrite vector
$\NPn^a$ given by~\eqref{eq:vector_n_general}
for our choice of constants of motion:
\begin{align}
    \begin{split}
        \boldsymbol{\NPn} = \frac{1}{\KNSigma} &\biggr[
        \left(\frac{\KNSigmaZ \CarterK}{\KKr + \KNSigmaZ} - \KNa^2 \sin^2\KNtheta\right) \pd_\KNv
        - \KKr\,  \pd_\KNr \\
               &\qquad{}+ \etheta\KKtheta\, \pd_\KNtheta
           + \left(\frac{\KNa \CarterK}{\KKr + \KNSigmaZ} - \KNa\right) \pd_\KNphi \biggr] \,,
    \end{split}
\end{align}
where we applied
\begin{equation}
    \frac{\KKr - \KNSigmaZ}{\KNDelta} = - \frac{\CarterK}{\KKr + \KNSigmaZ} \,,
\end{equation}
to make the expression explicitly regular on the horizon.

The parallel-propagated vectors given by \eqref{eq:KY_vectors} read
\begin{widetext}
\begingroup
\allowdisplaybreaks
\begin{subequations}
\begin{align}
    \begin{split}
        \boldsymbol{\NPl_\PP} &= \frac{1}{2\CarterK \KNSigma} \biggr[\!
        \left(\frac{\KNSigmaZ \CarterK \left(\Kr + \KNrho\right)\left(\Kr + \cconj{\KNrho}\right)}{\KKr + \KNSigmaZ} + \KNa^2 \sin^2\KNtheta \left(\KNSigma - \Kr^2 + 2\Kr\, \etheta\KKtheta \cot\KNtheta\right)\!\right) \pd_\KNv
        - \Big(\KKr \left(\KNSigma + \Kr^2\right) - 2 \KNr \Kr \KNSigmaZ\Big)\, \pd_\KNr \\[1.5ex]%
        &\phantom{
            \frac{1}{2\CarterK \KNSigma} \biggr[
        }\qquad{}- \Big(\etheta\KKtheta \left(\KNSigma - \Kr^2\right) - \KNa^2 \Kr \sin2\KNtheta\Big)\, \pd_\KNtheta
             + \KNa \left(
             \frac{\CarterK \left(\Kr+\KNrho\right)\left(\Kr+\cconj{\KNrho}\right)}{\KKr + \KNSigmaZ}
             + \KNSigma - \Kr^2 + 2\Kr\, \etheta\KKtheta \cot\KNtheta
         \right) \pd_\KNphi
        \biggr] \,,
    \end{split}
    \\[3.0ex]%
    \begin{split}
        \boldsymbol{\NPm_\PP} &= 
            -\frac{\ii}{\sqrt{2\CarterK}\, \KNSigma}\biggr[\!
            \left(\frac{\KNSigmaZ \CarterK \left(\Kr + \cconj{\KNrho}\right)}
            {\KKr + \KNSigmaZ}
            - \KNa^2 \Kr \sin^2 \KNtheta
            + \ii \KNa \cconj{\KNrho}\, \etheta\KKtheta \sin\KNtheta
             \right) \pd_\KNv
            - \left(\Kr\, \KKr - \cconj{\KNrho} \KNSigmaZ\right) \pd_\KNr \\[1.5ex]%
            &\phantom{
                - \frac{\ii}{\sqrt{2\CarterK}\, \KNSigma} \biggr[
            }\qquad{}+ \left(\Kr\, \etheta\KKtheta + \ii \KNa \cconj{\KNrho}\, \sin\KNtheta\right) \pd_\KNtheta 
            + \left(\frac{\KNa \CarterK \left(\Kr + \cconj{\KNrho}\right)}
               {\KKr + \KNSigmaZ}
               - \KNa \Kr
               + \frac{\ii\cconj{\KNrho}\, \etheta\KKtheta}
                {\sin\KNtheta}\right) \pd_\KNphi\biggr] \,.        
    \end{split}
\end{align}%
\label{eq:PP_tetrad}%
\end{subequations}%
\endgroup
\end{widetext}
Notice that the tetrad~\eqref{eq:PP_tetrad} has no well-defined limit as $\KNa \to 0$.
This is expected since the approach of~\cite{Kubiznak-2009}, which we used to
construct the frame, requires a non-degenerate principal conformal
Killing--Yano tensor --- a condition not satisfied by the Schwarzschild metric.
This behavior will be corrected at the end of this section in Eq.%
~\eqref{eq:rotated_tetrad} by a constant Lorentz transformation. 
Nevertheless, we postpone this trivial operation and first discuss the properties
of this intermediate tetrad with a more compact form.
Using the Kinnersley tetrad, the parallel-propagated one can be neatly
expressed as
\begingroup
\newcommand{\sepsize}{3.0ex}
\newcommand{\hspacemod}{\hspace{-1.5ex}}
\newcommand{\vspacemod}{\vspace{5ex}}
\begin{widetext}
\newcommand{\rightvect}{
    \begin{array}{c}
        \frac{\CarterK\KNSigmaZ}{\left(\KKr+\KNSigmaZ\right)\KNSigma}\, \boldsymbol{\NPl}_\KIN \\[\sepsize]
        \frac{\KKr+\KNSigmaZ}{2\KNSigmaZ}\, \boldsymbol{\NPn}_\KIN \\[\sepsize]
        \frac{\etheta \KKtheta + \ii \KNa\sin\KNtheta}{\sqrt{2}\,\KNrho}\, \boldsymbol{\NPm}_\KIN \\[\sepsize]
        \frac{\etheta \KKtheta - \ii \KNa\sin\KNtheta}{\sqrt{2}\,\cconj{\KNrho}}\, \cconj{\boldsymbol{\NPm}}_\KIN
    \end{array}
}
\begin{align}
    \begin{array}{ccccccccccc}
        \multirow[t]{4}{*}{\raisebox{-10.8ex}{$\left(\vphantom{\rightvect}\right.$\hspacemod}} &
        \boldsymbol{\NPl}_\PP &
        \multirow[t]{4}{*}{\raisebox{-10.8ex}{\hspacemod$\left.\vphantom{\rightvect}\right) = \left(\vphantom{\rightvect}\right.$\hspacemod}} &
        \dfrac{(\Kr+\KNrho)(\Kr+\cconj{\KNrho})}{2\CarterK} &
        \dfrac{(\Kr-\KNrho)(\Kr-\cconj{\KNrho})}{2\CarterK} &
        \dfrac{(\Kr-\KNrho)(\Kr+\cconj{\KNrho})}{2\CarterK} &
        \dfrac{(\Kr+\KNrho)(\Kr-\cconj{\KNrho})}{2\CarterK} &
        \multirow[t]{4}{*}{\raisebox{-10.8ex}{\hspacemod$\left.\vphantom{\rightvect}\right) \cdot \left(\vphantom{\rightvect}\right.$\hspacemod}} &
        \frac{\CarterK\KNSigmaZ}{\left(\KKr+\KNSigmaZ\right)\KNSigma}\, \boldsymbol{\NPl}_\KIN &
        \multirow[t]{4}{*}{\raisebox{-10.8ex}{\hspacemod$\left.\vphantom{\rightvect}\right)$}}
        \\[\sepsize]
        &\boldsymbol{\NPn}_\PP &&
        1 & 1 & 1 & 1 &&
        \frac{\KKr+\KNSigmaZ}{2\KNSigmaZ}\, \boldsymbol{\NPn}_\KIN &
        \\[\sepsize]
        &\boldsymbol{\NPm}_\PP &&
        -\ii\dfrac{\Kr+\cconj{\KNrho}}{\sqrt{2\CarterK}} &
        -\ii\dfrac{\Kr-\cconj{\KNrho}}{\sqrt{2\CarterK}} &
        -\ii\dfrac{\Kr+\cconj{\KNrho}}{\sqrt{2\CarterK}} &
        -\ii\dfrac{\Kr-\cconj{\KNrho}}{\sqrt{2\CarterK}} &&
        \frac{\etheta \KKtheta + \ii \KNa\sin\KNtheta}{\sqrt{2}\,\cconj{\KNrho}}\, \boldsymbol{\NPm}_\KIN &
        \\[\sepsize]
        &\cconj{\boldsymbol{\NPm}}_\PP &&
        \ii\dfrac{\Kr+\KNrho}{\sqrt{2\CarterK}} &
        \ii\dfrac{\Kr-\KNrho}{\sqrt{2\CarterK}} &
        \ii\dfrac{\Kr-\KNrho}{\sqrt{2\CarterK}} &
        \ii\dfrac{\Kr+\KNrho}{\sqrt{2\CarterK}} &&
        \frac{\etheta \KKtheta - \ii \KNa\sin\KNtheta}{\sqrt{2}\,\KNrho}\, \cconj{\boldsymbol{\NPm}}_\KIN &
    \end{array}
    \label{eq:wideTetradRelation}
\end{align}
\end{widetext}
\endgroup
The relation of the $\PP$ tetrad to the Kinnersley tetrad 
can be written as a composition of four elementary Lorentz transformations:
\begin{enumerate}
    \item a boost with parameter $A_{\KIN\to\PPm}$,
    \item a rotation about the new $\NPl^a_\PPm$ with parameter $c_{\PPm\to{\PPmm}}$,
    \item a rotation about the newest $\NPn^a_{\PPmm}$ with parameter $d_{{\PPmm}\to\PPmmm}$,
    \item a spin with parameter $\chi_{{\PPmmm}\to\PP}$.
\end{enumerate}
Note that these transformations do not commute, and their order is not fixed.
The order we suggest is based on the fact that we first set the size of the
vectors $\NPl^a$ and $\NPn^a$, then the direction of $\NPn^a$ which is set
by the condition that it is non-twisting. The direction of vector $\NPl^a$
is then fixed, and the spin, which is the only one with some freedom (we will
discuss that in Sec.~\ref{sec:axial_IH}), is performed last. Additional
comments on the transformations are made in Sec.~\ref{sec:slow_rotation_lorentz}.
For the presented order of transformations, 
the parameters, as determined from the algebraic relation
\eqref{eq:wideTetradRelation}, are
\begin{align}
    A^2_{\KIN\to\PPm} &= \frac{2\KNSigmaZ}{\KKr + \KNSigmaZ} \,, \label{eq:boost_gen}\\[1ex]%
    c_{\PPm\to\PPmm} &= \frac{\etheta\KKtheta + \ii\KNa\sin\KNtheta}{\sqrt{2}\cconj{\KNrho}} \,, \label{eq:rot_l_gen}\\[1ex]%
    d_{\PPmm\to\PPmmm} &= \frac{\Kr - \cconj{\KNrho}}{\sqrt{2}\, \left(\etheta\KKtheta + \ii\KNa\sin\KNtheta\right)} \,, \label{eq:rot_n_gen}\\[1ex]%
    \eu^{2\ii\chi_{\PPmmm\to\PP}} &= -\ii\, 
    \frac{\etheta\KKtheta - \ii\KNa\sin\KNtheta}{\sqrt{\CarterK}} \,.
\end{align}
It is simple to verify that $A_{\KIN\to\Hm}$ and $A_{\KIN\to\PPm}$ as well as
$c_{\Hm\to\Hmm}$ and $c_{\PPm\to\PPmm}$
correspond to each other on the horizon, recall Eqs.
\eqref{eq:boost_hor} and \eqref{eq:rot_l_hor} respectively.
On the horizon, the parameter $\chi_{\PPmmm\to\PP}$ vanishes
for the choice $\CarterK \horeq \KNa^2\sin^2\KNtheta$.

The Newman--Penrose projections of the Weyl tensor onto the parallel-propagated
tetrad~\eqref{eq:PP_tetrad} are given by
\begin{subequations}
\begin{alignat}{2}
    \NPPsi^\PP_0 &={}& - \,\frac{1}{2\CarterK}\, &\frac{3\NPPsi^\KIN_2}{2\,\cconj{\KNrho}^2} \left(\Kr^2-\cconj{\KNrho}^2\right)^2,\\[0.5ex]
    \NPPsi^\PP_1 &={}& - \frac{\ii}{\sqrt{2\CarterK}}\, &\frac{3\NPPsi^\KIN_2}{2\,\cconj{\KNrho}^2} \left(\Kr^2-\cconj{\KNrho}^2\right) \Kr \,,\\[0.5ex]
    \NPPsi^\PP_2 &={}&  &\frac{\NPPsi^\KIN_2}{2\,\cconj{\KNrho}^2} \left(3\Kr^2-\cconj{\KNrho}^2\right),\\[0.5ex]
    \NPPsi^\PP_3 &={}& \ii \,\sqrt{2\CarterK}\, &\frac{3\NPPsi^\KIN_2}{2\,\cconj{\KNrho}^2} \,\Kr\,,\\[0.5ex]
    \NPPsi^\PP_4 &={}& -\,2\CarterK\, &\frac{3\NPPsi^\KIN_2}{2\,\cconj{\KNrho}^2} \,.
\end{alignat}
\label{eq:PP_tetrad_Psi}%
\end{subequations}%
Recall that the only non-zero Kinnersley projection is 
\begin{equation}
    \NPPsi^\KIN_2 = - \frac{\massBH }{ \cconj{\KNrho}^3} \,.
\end{equation}

Unlike the Weyl scalars, the spin coefficients do not seem to have
a simple representation. Both computing them directly from the $\PP$ tetrad
or using the knowledge of Lorentz transformations
\eqref{eq:boost_gen}-\eqref{eq:rot_n_gen} and the spin coefficients of the
Kinnersley tetrad result in lengthy expressions.
Let us focus on the horizon initial data instead. Moreover, we employ the choice
$\CarterK \horeq \KNa^2\sin^2\KNtheta$ to simplify them further.
We find that the non-vanishing spin coefficients on the horizon are as follows
\begingroup
\allowdisplaybreaks
\begin{subequations}
\begin{align}
    \Smu_\PP &\horeq - \frac{\KNrp + \massBH}{2\KNSigma} - \frac{\KNrp - \KNrm}{4\KNSigmaZ}
    \,, \\
    \Slambda_\PP &\horeq - \frac{\KNa^2 \sin^2\KNtheta}{\cconj{\KNrho}^3} - \frac{\left(\KNrp - \KNrm\right)\KNa^2 \sin^2\KNtheta}{8\massBH \KNrp \cconj{\KNrho}^2}
    \,, \\
    \Spi_\PP &\horeq \frac{\ii\,\KNrp\,\csc\KNtheta}{\sqrt{2}\, \KNa \cconj{\KNrho}}
    \,, \\
    \Sa_\PP &\horeq - \frac{\ii\,\csc\KNtheta}{\sqrt{2}\, \KNa}
    \,, \\
    \Salpha_\PP &\horeq - \frac{\cot\KNtheta}{2\sqrt{2}\, \cconj{\KNrho}}
    \,, \\
    \Sbeta_\PP &\horeq - \frac{\ii\,\csc\KNtheta}{2\sqrt{2}\, \KNa} - \frac{\ii\,\KNrp\,\csc\KNtheta}{2\sqrt{2}\,\KNa \KNrho}
    \,, \\
    \Sepsilon_\PP &\horeq \frac{\massBH \KNrp \left(\KNSigma - 2 \KNrp \cconj{\KNrho}\right)}{\cconj{\KNrho}^3\, \KNa^2 \sin^2\KNtheta}
    \,, \\
    \Skappa_\PP &\horeq - \frac{\ii\left(\KNrho + 2\KNrm\right)}{2\sqrt{2}\,\KNrm \KNa \sin^3\KNtheta}
    \,, \\
    \Srho_\PP &\horeq \frac{\KNrm-\KNrp}{16\massBH \KNrp}
    \,, \\
        \Ssigma_\PP &\horeq \frac{\KNSigma + 2\massBH\KNrp}
        {2\KNrho\KNa^2\sin^2\KNtheta}
        - \frac{\cconj{\KNrho}\!\left(\KNSigma + 6\massBH\KNrp + \KNrmSQ\sin^2\KNtheta\right)}
        {16\massBH\KNrho\KNa^2\sin^2\KNtheta}
    \,.
\end{align}
\end{subequations}
\endgroup

In the next step, we still need to use the Lorentz transformation to obtain the
desired vectors. Importantly, unlike in~\eqref{eq:Kofron-2024_tetrad},
this transformation is coordinate independent due to parallel propagation.
Therefore, it is sufficient to find the transformation
on the horizon. Conveniently, we are already equipped with the
expressions~\eqref{eq:tetrad_horizon}. Thus, we can evaluate
the vectors $\NPl^a_\PP$ and $\NPm^a_\PP$
on the horizon and compare them to determine the constant Lorentz transformation
that yields the proper tetrad everywhere.

On the horizon, we set $\Kr \horeq 0$ as usual and find that
we need to perform a rotation about $\NPn^a$ with the parameter%
\footnote{
    Notice that $\KNa^2 - \CarterK \geq 0$.
}
\begin{equation}
    d =  - \ii\, \frac{\KNrp - \ii \KNa \cos\KNthetaZ}{\sqrt{2\CarterK}} 
    \overset{\CarterK = \KNa^2 \sin\KNthetaZ}{=}
    - \frac{\ii\,\KNrp + \ethetaD \sqrt{\KNa^2 - \CarterK}}{\sqrt{2\CarterK}}
    \,,
    \label{eq:constant_n_rotation}
\end{equation}
which yields the correct vector $\NPl^a$. 
Generally, we might also need a spin to set the vectors $\NPm^a$ and
$\cconj{\NPm}^a$ identical to \eqref{eq:tetrad_horizon},
nevertheless, \eqref{eq:PP_tetrad} is such that no spin is needed.

\begin{subequations}
\textbf{The final tetrad} is given by 
\begin{align}
    \NPl^a &= \NPl^a_\PP + \cconj{d}\, \NPm^a_\PP + d\, \cconj{\NPm}^a_\PP
    + \left|d\right|^2 \NPn^a \,, \\[1ex]
    \NPm^a &= \NPm^a_\PP + d\, \NPn^a .
\end{align}%
\label{eq:rotated_tetrad}%
\end{subequations}%
Fixing the tetrad on the horizon also fixes the gauge freedom,
which recovers the limit $\KNa\to0$.

We already possess some of the initial data for the isolated horizon,
since they are unaffected by the null
rotation about $\NPn^a$: $\Smu$, $\Slambda$, $\NPPsi_4$.
To obtain the complete set, we need to supplement the spin coefficients
$\Spi$, $\Sa$, and $\Sepsilon$. The full set is:
\begingroup
\allowdisplaybreaks
\begin{subequations}
\begin{align}
    \Smu &= \Smu_\PP \horeq - \frac{\KNrp + \massBH}{2\KNSigma} - \frac{\KNrp - \KNrm}{4\KNSigmaZ}
    \,, \\
    \Slambda &= \Slambda_\PP \horeq - \frac{\KNa^2 \sin^2\KNtheta}{\cconj{\KNrho}^3} - \frac{\left(\KNrp - \KNrm\right)\KNa^2 \sin^2\KNtheta}{8\massBH \KNrp \cconj{\KNrho}^2}
    \,, \\
    \Spi &\horeq \frac{\left(2\ii\KNa\left(\KNrp+\massBH\right) + \KNrm\left(\KNrp -\KNrm\right) \cos\KNtheta\right)\sin\KNtheta}{4\sqrt{2}\,\massBH \cconj{\KNrho}^2}
    \,, \\
    \Sa &\horeq \frac{\ii \KNa - \KNrp \cos\KNtheta}{\sqrt{2}\,\cconj{\KNrho}^2 \sin\KNtheta}
    \,, \\
    \Sepsilon &\horeq \frac{\KNrp - \KNrm}{4\KNSigmaZ}
    \,,
\end{align}
\end{subequations}
\endgroup
on $\SS_0$ and
\begin{align}
    \NPPsi_4 = \NPPsi^\PP_4 = \frac{3\massBH\CarterK}{\cconj{\KNrho}^5} \,,
\end{align}
on $\NN_0$.
It is clear that $\Smu$ is real, as it should be for non-twisting congruence,
and that $\Sepsilon$ gives the correct surface gravity, using 
$\surfkappa_{(\NPl)} = \Sepsilon + \cconj{\Sepsilon}$. Recall that the spin
coefficient $\Smu$ gives the expansion of $\NPn^a$ ($2\Re\Smu =
\OSexp_{\NPn}$), $\Slambda$ gives its shear, and $\Sa$ is the connection of
$\SS_0$. The spin coefficients $\Salpha$ and $\Sbeta$ can be obtained from
$\Spi$ and $\Sa$ using
\begin{align}
    \Spi &= \Salpha + \cconj{\Sbeta} \,, &
    \Sa  &= \Salpha - \cconj{\Sbeta} \,,
\end{align}
where the first equality is a consequence of the gauge chosen
by~\cite[Eq. (34)]{Krishnan-2012} for an isolated horizon, which holds for the
tetrad~\eqref{eq:rotated_tetrad}, and the
second is the definition of $\Sa$, usually introduced under such a gauge.

\add{
Note that due to the properties of the isolated horizon and the formulation of
the initial value problem \cite{Krishnan-2012},
it holds that $\Srho \horeq \Ssigma \horeq \Skappa \horeq 0$
and $\Stau = \Snu = \Slambda = 0$.
On the horizon, we also have $\NPPsi_0 \horeq \NPPsi_1 \horeq 0$.
}

Moreover, the initial data include the angular components of the vector
$\NPm^a$ on the horizon.
However, these depend on a spin transformation that we could
optionally perform since there is freedom in the choice
of the vectors $\NPm^a$ and $\cconj{\NPm}^a$ of the Krishnan tetrad.
We will show a preferred choice of such a transformation and the resulting
components of $\NPm^a$ in Sec.~\ref{sec:axial_IH}. 

\begin{figure}
    \centering
    \figinput{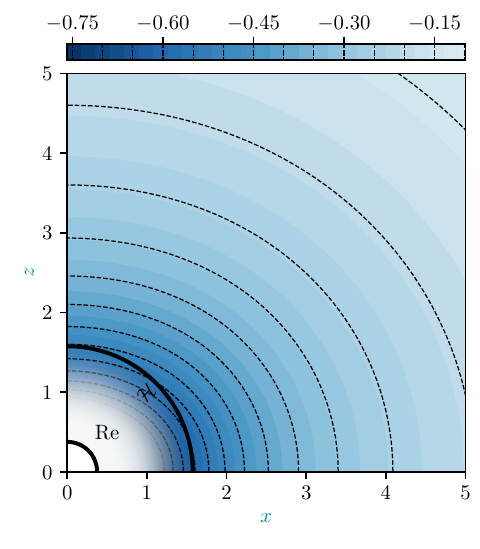}
    \caption{
        Spin coefficient $\Smu$ given by tetrad~\eqref{eq:rotated_tetrad}
        for black hole with $\massBH = 1$ and $\KNa = 0.8$ in the Cartesian
        version of ingoing null coordinates~\eqref{eq:cartesin_kerr_null_coor}.
        The imaginary part is zero (as it should be for a non-twisting geodesic
        congruence) and the displayed real part is the expansion of the
        congruence (in the direction of $\NPn^a$).
    }
    \label{fig:PPcr_mu}
\end{figure}

\begin{figure}
    \centering
    \figinput{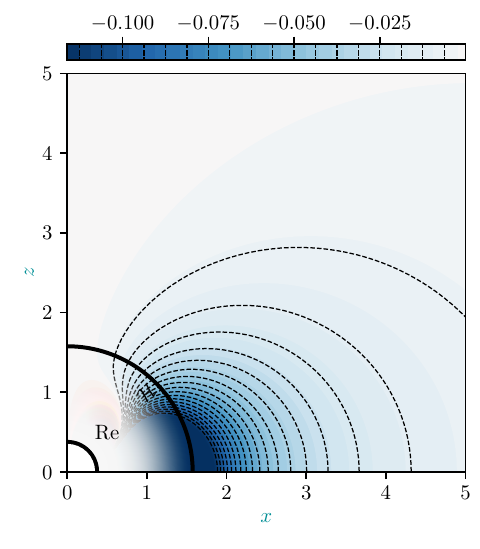}
    \caption{
        Spin coefficient $\Slambda$ given by tetrad~\eqref{eq:rotated_tetrad}
        for black hole with $\massBH = 1$ and $\KNa = 0.8$ in the Cartesian
        version of ingoing null coordinates~\eqref{eq:cartesin_kerr_null_coor}.
        This spin coefficient represents the shear of the congruence (in the
        direction of $\NPn^a$). The plot shows the real part,
        the imaginary part is zero.
    }
    \label{fig:PPcr_lambda}
\end{figure}

\begin{figure*}
    \centering
    \figinput{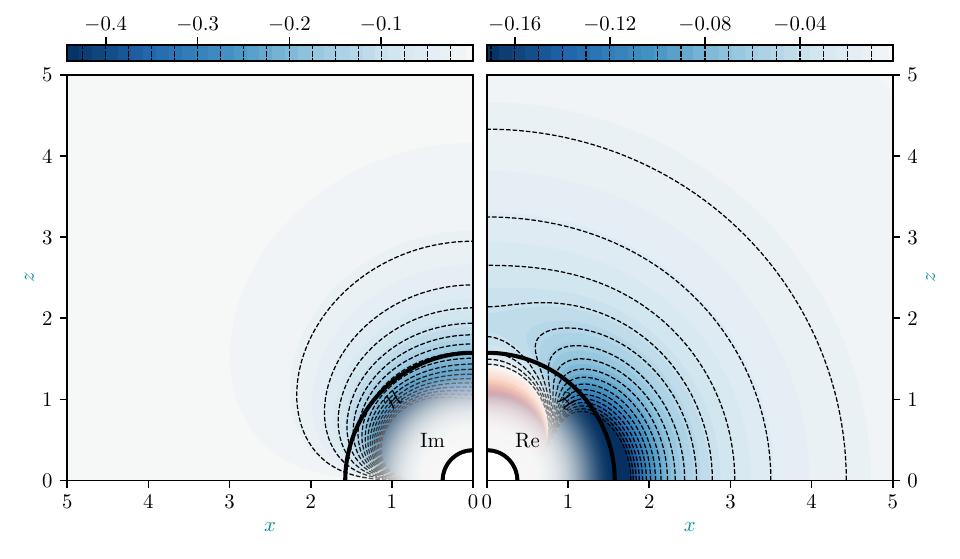}
    \caption{
        Weyl scalar $\NPPsi_2$ given by tetrad~\eqref{eq:rotated_tetrad}
        for black hole with $\massBH = 1$ and $\KNa = 0.8$ in the Cartesian
        version of ingoing null coordinates~\eqref{eq:cartesin_kerr_null_coor}.
        The real (right) and imaginary (left) parts are plotted separately
        for the same quadrant ($\KNtheta \in \langle0,\Cpi/2\rangle$), the
        imaginary part is mirrored in the plot.
    }
    \label{fig:PPcr_Psi2}
\end{figure*}

\begin{figure*}
    \centering
    \figinput{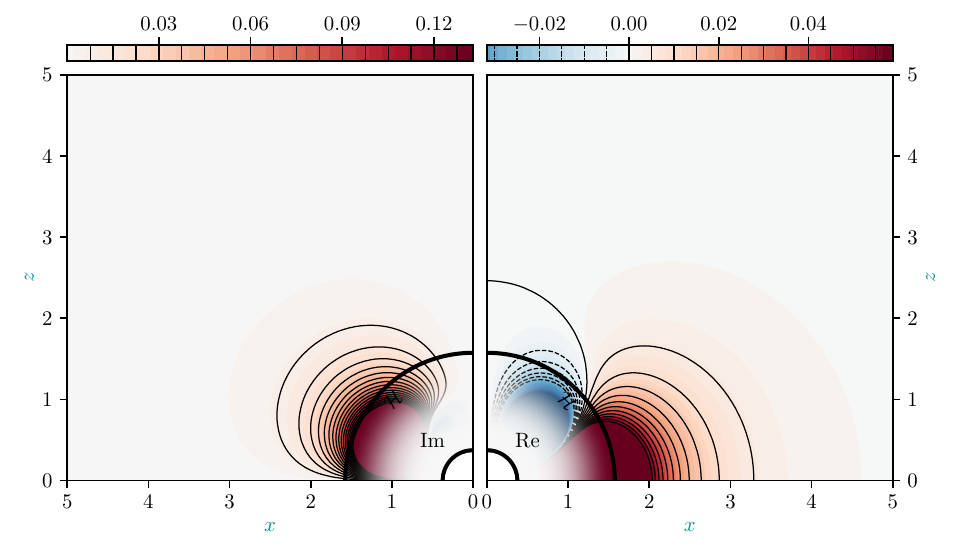}
    \caption{
        Weyl scalar $\NPPsi_4$ given by tetrad~\eqref{eq:rotated_tetrad}
        for black hole with $\massBH = 1$ and $\KNa = 0.8$ in the Cartesian
        version of ingoing null coordinates~\eqref{eq:cartesin_kerr_null_coor}.
        The real (right) and imaginary (left) parts are plotted separately
        for the same quadrant ($\KNtheta \in \langle0,\Cpi/2\rangle$), the
        imaginary part is mirrored in the plot.
    }
    \label{fig:PPcr_Psi4}
\end{figure*}

In Figs.~\ref{fig:PPcr_mu}--\ref{fig:PPcr_Psi4}, the most important 
Newman--Penrose scalars are shown. They are plotted in Cartesian coordinates
given by
\begin{align}
    \KNEx &= \KNr \sin\KNtheta \,, &
    \KNEz &= \KNr \cos\KNtheta \,.
    \label{eq:cartesin_kerr_null_coor}
\end{align}
The area inside the outer horizon is intentionally faded toward the inner
horizon for two reasons. First, the focus is on the physically relevant outer region, and
the precision towards the inner horizon may deteriorate. Second,
the quantities typically grow large in magnitude there, and including
them in the contour range would obscure the behavior outside the outer
horizon. Nevertheless, the extension inside the outer horizon remains valid, and
its qualitative character is visible in the figures.


\section{Tetrad in the coordinates adapted to the isolated horizons}\label{sec:coordinates}
While the tetrad~\eqref{eq:rotated_tetrad} does satisfy all the geometrical
properties needed to describe an isolated horizon, it is not yet represented
in coordinates adapted to the isolated horizon. Such coordinates were introduced
in \cite{Krishnan-2012}; we will denote them: ($\Kv$, $\Kr$, $\Ktheta$, $\Kphi$).

We have, nevertheless, already met two of them. The radial coordinate $\Kr$ is
defined to be the affine parameter that helped us ensure parallel propagation
of the tetrad. On the other hand, the polar angle $\Ktheta$ is defined to be
constant along each geodesic but otherwise varies in the manifold, just like
the Carter constant $\CarterK$, to which it is connected by
Eq.~\eqref{eq:K_is_a2_sin2_theta}, where we identified $\KNthetaZ = \Ktheta$. Hence, we have
$\CarterK = \KNa^2\sin^2\Ktheta$.

Let us start by writing differentials for the new coordinates:
\begin{subequations}
\begin{alignat}{5}
    \d\Kv               &={} & \Kv_{,\KNv}\,&\d\KNv +{}& \Kv_{,\KNr}\, &\d\KNr +{}& \Kv_{,\KNtheta}\, &\d\KNtheta \,, \\
    \d \Kr              &={} & && \Kr_{,\KNr}\,     &\d\KNr +{}& \Kr_{,\KNtheta}\,     &\d\KNtheta\,,\\
    \d \Ktheta          &={} & && \Ktheta_{,\KNr}\, &\d\KNr +{}& \Ktheta_{,\KNtheta}\, &\d\KNtheta\,,\\
    \d \tilde{\Kphi}    &={} & && \Jphi_{,\KNr}\,   &\d\KNr +{}& \Jphi_{,\KNtheta}\,   &\d\KNtheta{} +{} \d\KNphi\,,
    \label{eq:transform_phi}
\end{alignat}%
\label{eq:coordinate_transform}%
\end{subequations}%
where we decorated the coordinate $\Kphi$ with a tilde to indicate that
it will undergo a slight modification later, yielding the final coordinate $\Kv$.
The coefficients for $\d\Kv$ are given by $\d\Kv \equiv \NPn_a \d x^a$ so that
\begin{equation}
    \d\Kv = \d\KNv - \frac{\CarterK}{\KKr + \KNSigmaZ}\, \d\KNr
        - \etheta\KKtheta\, \d\KNtheta \,.
\end{equation}

By construction, see \cite{Krishnan-2012}, coordinates $\Kv$, $\Ktheta$, and
$\Kphi$ are parallel transported off the horizon:
\begin{align}
    \NPn^a \nabla_a \Kv &= 0 \,,&
    \NPn^a \nabla_a \Ktheta &= 0 \,,&
    \NPn^a \nabla_a \Kphi &= 0 \,,
\end{align}
and we require the same also for $\tilde{\Kphi}$.
While $\Kv$ and $\Ktheta$ (by means of $\CarterK$) already satisfy this condition,
we need to require
\begin{align}
    -\KKr\, \Jphi_{,\KNr} + \etheta\KKtheta\, \Jphi_{,\KNtheta}
    +\frac{\KNa \CarterK}{\KKr + \KNSigmaZ} - \KNa
    &=0 \,,
    \label{eq:J_function}
\end{align}
for $\tilde{\Kphi}$ to be adapted to isolated horizons. This relation, in turn, fixes
the $(\KNr, \KNtheta)$ dependence of $\Jphi$.
Note that the new coordinates are well defined on the horizon.

By inverting relations~\eqref{eq:coordinate_transform} and transforming
the metric~\eqref{eq:Kerr_null_coordinates} into the coordinates
($\Kv$, $\Kr$, $\Ktheta$, $\tilde{\Kphi}$), we arrive at
\begin{widetext}
\newcommand{\upsilonterm}{\left(\KNSigma \KNSigmaZ + 2\massBH\KNa^2\,\KNr \sin^2\KNtheta\right)}
\begin{align}
\begin{split}
    \dss &= 
    \left(1 - \frac{2\massBH \KNr}{\KNSigma}\right) \d \Kv^2
    - 2 \left(\d \Kr - \frac{\Kr_{,\KNtheta} \KNSigma + \left(\KNSigma - 2\massBH\KNr\right) \etheta\KKtheta + 2\massBH\KNa\,\KNr\, \Jphi_{,\KNtheta} \sin^2\KNtheta}{\Ktheta_{,\KNtheta} \KNSigma}\, \d \Ktheta - \frac{2\massBH\KNa\,\KNr\sin^2\KNtheta}{\KNSigma}\, \d \tilde{\Kphi}\right)\d \Kv \\[1ex]
    &\qquad{}
- \frac{\KNSigma^2 -\left(\KNSigma - 2\massBH\KNr\right)\KKtheta^2 + 4 \massBH \KNa\,\KNr\, \Jphi_{,\KNtheta}\, \etheta\KKtheta\sin^2\KNtheta + {{\Jphi_{,\KNtheta}}{}}^2\upsilonterm\sin^2\KNtheta}{{\Ktheta_{,\KNtheta}}^2 \KNSigma}\, \d \Ktheta^2 \\[1ex]
    &\qquad{}
    + 2\sin^2\KNtheta\, \frac{\Jphi_{,\KNtheta}\upsilonterm + 2\massBH\KNa\,\KNr\,\etheta\KKtheta}{\Ktheta_{,\KNtheta} \KNSigma}\, \d \Ktheta\, \d \tilde{\Kphi}
    - \frac{\upsilonterm \sin^2\KNtheta}{\KNSigma}\, \d \tilde{\Kphi}^2 \,.
\end{split}
\end{align}
It is worth
observing that while the metric is quite complicated, its determinant is
very simple:
\begin{align}
    \det g &= -\frac{\KKr^2\sin^2\KNtheta}{\Ktheta_{,\KNtheta}{}^2} \,.
\end{align}

The parallel-propagated tetrad~\eqref{eq:PP_tetrad} takes in these coordinates
the form
\begin{subequations}
\begin{align}
\begin{split}
    \boldsymbol{\NPl}_\PP &= \pd_\Kv
        + \frac{1}{2\CarterK\, \KKr \KNSigma} \biggr[
         \Big(-\KKr \KNSigma \left(\KNSigma + 2 \etheta\KKtheta\, \Kr_{,\KNtheta} + \Kr^2\right) + 2 \KNr \Kr \KNSigmaZ \left(\KNSigma + \etheta\KKtheta\,\Kr_{,\KNtheta}\right) + \KNa^2 \KKr \Kr \Kr_{,\KNtheta} \sin(2\KNtheta)\Big)\, \pd_\Kr \\[0.5ex]
                          &\qquad{}- \Ktheta_{,\KNtheta}\Big(2\etheta\KKtheta \left(\KKr\KNSigma - \KNr\Kr\KNSigmaZ\right) - \KNa^2\Kr\,\KKr\sin(2\KNtheta)\Big)\, \pd_\Ktheta \\[0.5ex]
        &\qquad{}+ 2\Big(\left(\KKr \KNSigma - \KNr\Kr\KNSigmaZ\right) \left(\KNa - \etheta\KKtheta\Jphi_{,\KNtheta}\right) + \KNa\Kr \big(\CarterK \KNr + \KKr \cot\KNtheta \left(\etheta\KKtheta + \KNa\, \Jphi_{,\KNtheta} \sin^2\KNtheta\right)\!\big)\Big)\, \pd_{\tilde{\Kphi}}
    \biggr] \,,
\end{split}
\\[1.5ex]
\begin{split}
    \boldsymbol{\NPn}_\PP &= - \pd_\Kr \,,
\end{split}
\\[1.5ex]
\begin{split}
    \boldsymbol{\NPm}_\PP &= \frac{1}{\sqrt{2\CarterK}\, \KKr \KNrho} \biggr[
        \Big(\KNa \KKr \Kr_{,\KNtheta} \sin\KNtheta + \ii \big(\KKr \Kr \KNrho - \left(\KNSigma + \etheta\KKtheta \Kr_{,\KNtheta}\right) \KNSigmaZ\big)\Big)\, \pd_\Kr
        + \Ktheta_{,\KNtheta} \Big(\KNa \KKr \sin\KNtheta - \ii\, \etheta\KKtheta \KNSigmaZ\Big)\, \pd_\Ktheta \\[0.5ex]
                      &\qquad{}+ \Big(\KKr\csc\KNtheta\left(\etheta\KKtheta + \KNa \Jphi_{,\KNtheta} \sin^2\KNtheta\right) - \ii \big(\KNa \CarterK - \KNSigmaZ \left(\KNa - \etheta\KKtheta\, \Jphi_{,\KNtheta}\right)\!\big)\!\Big)\, \pd_{\tilde{\Kphi}}
    \biggr] \,.
\end{split}
\end{align}
\label{eq:PP_tetrad_transformed}
\end{subequations}
\end{widetext}

In order to obtain the final tetrad, we not only need the constant null rotation
about $\NPn^a$, recall~\eqref{eq:rotated_tetrad}.
Unfortunately, we also need one additional coordinate transformation
-- a rigid rotation with the angular velocity of the horizon.%
\footnote{Note that neither of the two transformations changes the vector $\NPn^a_\PP$.}
Namely,
\begin{equation}
    \Kphi = -\frac{\KNa}{\KNrp\!^2 + \KNa^2}\,\Kv + \tilde{\Kphi} \,,
\end{equation}
where the factor in front of $\d\Kv$
is chosen so that the $\pd_\Kphi$ component of $\NPl^a$ vanishes on the horizon,
recall
Eq.~\eqref{eq:l_on_H}.
While the transformation itself is trivial, the resulting expressions are
substantially lengthier. For that reason, we have chosen not to include it
in~\eqref{eq:transform_phi} and to show
the metric and tetrad without it. 
We leave this trivial addition to the kind reader.
It is worth noting that since both $\Kv$ and $\tilde{\Kphi}$ are
parallel propagated by construction and the factor is constant,
$\Kphi$ is also parallel propagated, as it should be.

In the coordinates adapted to the isolated horizon, the tetrad has the 
special form
\begin{subequations}
    \begin{align}
        \boldsymbol{\NPl} &= \pd_\Kv + \TU\,\pd_\Kr + {\TX^2}\,\pd_\Ktheta
            + \TX^3\,\pd_\Kphi \,, \\
        \boldsymbol{\NPn} &= -\,\pd_\Kr \,, \\
        \boldsymbol{\NPm} &= \TOmega\,\pd_\Kr + {\Txi^2}\,\pd_\Ktheta
            + \Txi^3\,\pd_\Kphi \,,
    \end{align}%
    \label{eq:general_tetrad}%
\end{subequations}%
and we need to find the metric functions $\TU$, $\TX^I$, $\TOmega$ and~$\Txi^I$. 
In Appendix~\ref{app:metric_functions}, relations connecting the (metric functions
of) tetrad~\eqref{eq:PP_tetrad_transformed} and tetrad~\eqref{eq:general_tetrad}
(with the spin applied) can be found.
These metric functions depend on the last gauge freedom in the choice of the
tetrad that we discuss in the following subsection.

\subsection{Axial isolated horizon}\label{sec:axial_IH}
In \cite{Ashtekar-2004}, a preferred choice
of tetrad for axially symmetric isolated horizons has been introduced.
We now employ this choice
to eliminate the remaining freedom in the choice of the angular
coordinates and the vectors $\NPm^a$ and~$\cconj{\NPm}^a$.

Ashtekar first proposed introducing coordinates $\Kzeta$, $\Kphi$
on $\SS$, the two-sphere foliation of the horizon,
such that the two-metric has the form
\begin{equation}
    \boldsymbol{q} = - \EuclidR^2 \left(h(\Kzeta)^{-1}\, \d \Kzeta^2
    + h(\Kzeta) \, \d \Kphi^2\right) ,
    \label{eq:metric_axi_ih}
\end{equation}
where $\EuclidR$ denotes the Euclidean radius defined by the area
of the two-sphere $\SS$, which is $\area = 4 \Cpi\EuclidR^2$.
The function $h(\Kzeta)$ must satisfy:
\begin{align}
    \begin{aligned}
        h(\pm 1) &= 0 \,, \\
        h'(\pm 1) &= \mp 2 \,,
    \end{aligned}
    \label{eq:axial_h_cond}
\end{align}
to avoid conical singularities.
Our coordinates are of this form if
\begin{equation}
    \Jphi_{,\KNtheta} \horeq 0 \,.
    \label{eq:Jphi_hor_cond}
\end{equation}
The Euclidean radius and function $h$ are given as
\begin{align}
    \EuclidR &= \sqrt{2\massBH\KNrp} \,, &
    h &= \frac{2\massBH\KNrp \sin^2\Ktheta}{\KNrp\!^2 + \KNa^2 \cos^2\Ktheta} \,.
\end{align}

Equipped with such a two-metric, there is a particularly nice choice of the
vectors $\NPm^a$ and $\cconj{\NPm}^a$ that span the two-spheres.
It is given by:%
\footnote{The signs are chosen based on the Kinnersley tetrad, which satisfies
such a property in the limit $\KNa\to0$.}
\begin{equation}
    \boldsymbol{\NPm}_\text{A} \horeq - \frac{1}{\sqrt{2}\,\EuclidR} \left(\sqrt{h}\,
    \pd_\Kzeta - \frac{\ii}{\sqrt{h}}\, \pd_\Kphi\right) .
    \label{eq:axial_m_vector}
\end{equation}

To achieve the canonical form for an axially symmetric
isolated horizon, we need to perform an additional \emph{constant} spin
transformation, 
with the coefficient $\chi_\text{A}$ given by
\begin{equation}
    \eu^{2\ii\chi_\text{A}} = \sqrt{\frac{\KNrp + \ii \KNa \cos\Ktheta}
    {\KNrp - \ii \KNa \cos\Ktheta}} \,\,,
\end{equation}
which does not break the parallel propagation.

\section{Coordinates adapted to the isolated horizons}\label{sec:particular_solutions}
So far, we have constructed a non-twisting parallel-propagated tetrad
that is adapted to the isolated horizon. Although we also found the transformation
to the coordinates adapted to the isolated horizon, the coordinates have not yet been
given explicitly and, consequently, neither has the tetrad.
Hence, we need to find the unknown functions
$\Kr$, $\Ktheta$ (given through $\CarterK$), and $\Kphi$.
They are governed by Eqs.%
~\eqref{eq:affine_parameter},
\eqref{eq:Carter_geodesic_cond}, and
\eqref{eq:J_function}:
\begin{align}
    \KKr\, \Kr_{,\KNr} - \etheta\KKtheta\, \Kr_{,\KNtheta} &= \KNSigma \,,
    \label{eq:equation_for_s} \\
    \KKr\,\CarterK_{,\KNr} - \etheta\KKtheta\,\CarterK_{,\KNtheta} &= 0 \,,
    \label{eq:equation_for_K} \\
    \KKr\, \Jphi_{,\KNr} - \etheta\KKtheta\, \Jphi_{,\KNtheta} &=
    \frac{\KNa \CarterK}{\KKr + \KNSigmaZ} - \KNa \,,
    \label{eq:equation_for_J}
\end{align}
with the boundary conditions on the horizon being $\Kr\horeq0$,
$\CarterK\horeq\KNa^2 \sin^2 \Ktheta$ and $\tilde\Kphi\horeq\KNphi$.

Let us start with the simplest case --- the axis of rotation where
$\KNtheta = 0$. Any choice of $\CarterK$ which leads to a finite
expansion of $\NPn^a$ on the horizon must satisfy $\CarterK(\KNrp, 0) = 0$.
From the rotational symmetry, it follows that
\begin{equation}
    \CarterK(\KNr, 0) = 0 \,.
\end{equation}
Using this solution in Eq.~\eqref{eq:equation_for_s}, we obtain
\begin{equation}
    \Kr(\KNr, 0) = \KNr - \KNrp \,.
    \label{eq:s_on_h}
\end{equation}
Similarly, for $\tilde{\Kphi}$, we get from Eq.~\eqref{eq:J_function}
\begin{equation}
    \tilde{\Kphi}(\KNr, 0, \KNphi) = \KNphi - \arctan\frac{\KNr}{\KNa}
    + \arctan\sqrt{\frac{\KNrp}{\KNrm}} \,.
\end{equation}

On the equator, where $\KNtheta = \tfrac{\Cpi}{2}$,
the situation is  more complicated.
By similar arguments, the Carter constant is:
\begin{equation}
    \CarterK(\KNr, \tfrac{\Cpi}{2}) = \KNa^2 \,.
\end{equation}
For $\Kr$ and $\tilde{\Kphi}$, we obtain the following equations:
\begin{align}
    \Kr_{,\KNr}(\KNr, \tfrac{\Cpi}{2}) &= \frac{\KNr^2}{\sqrt{\KNSigmaZ^2 - \KNa^2 \KNDelta}} \,, 
    \label{eq:s_eqs_equator} \\
    \tilde{\Kphi}_{,\KNr}(\KNr, \tfrac{\Cpi}{2}) &= -\frac{\KNa\left(\KNr^2 + \sqrt{\KNSigmaZ^2 - \KNa^2 \KNDelta}\right)}{\sqrt{\KNSigmaZ^2 - \KNa^2 \KNDelta}\left(\KNSigmaZ + \sqrt{\KNSigmaZ^2 - \KNa^2 \KNDelta}\right)} \,.
    \label{eq:phi_eqs_equator}
\end{align}
These equations can be solved (integrated) analytically, and they both
result in expressions involving elliptical integrals.
Nevertheless, the following section presents an approach employing elliptic
integrals, with integral of the right side of~\eqref{eq:s_eqs_equator} arising as a
special case. For Eq.~\eqref{eq:phi_eqs_equator}, the corresponding integral is
given in the Appendix, Eq~\eqref{eq:phi_eqs_equator_sol}.

\subsection{Analytical approach}\label{sec:analytical_solution}
It might be surprising
that it is possible to obtain analytical results.
Prescribing the initial data on the horizon, $\CarterK$ is being
parallel propagated along the geodesics
($\CarterK$ varying over the horizon yet constant along a particular geodesic).
For null geodesics on the Kerr
background, the problem is completely solved.  We may either use the approach
introduced in \cite{Cieslik-2023}, which utilizes the Weierstrass
$\mathscr{p}$ function and was deployed in \cite{Kofron-2024},
or one developed in \cite{Gralla-2020}, which takes advantage of
elliptic functions. 

Since we follow the latter approach, let us first clarify the notation:
we use the (incomplete) elliptic integrals of the first kind $F(\phi\,|\,m)$ as
well as the second kind $E(\phi\,|\,m)$. We also use Jacobi elliptic functions,
namely $\dn(\phi\,|\,m)$, and the Jacobi epsilon function $\mathcal{E}(\phi\,|\,m)$.
Finally, just like \cite{Gralla-2020}, we use exactly the convention of
\emph{Wolfram Mathematica}; for exact definitions, see \cite{NIST:DLMF}.

We are following the notation of \cite{Gralla-2020} and
kindly ask the reader to check this reference for all the definitions that
we only reference but do not retype in this paper.
We initially set $\energy=1$, so there
is no need to introduce rescaled quantities. The only difference is that 
\cite{Gralla-2020} uses a different definition of the Carter constant,
such that in our notation
$\CarterK=\KNa^2+\Cartereta$.

It is convenient to introduce the so-called Mino time $\minoT$,
\cite[Eq. (9)]{Gralla-2020},
\begin{equation}
    \frac{\d x^a}{\d \minoT} = \KNSigma \NPn^a \,, 
\end{equation}
which is measured from the horizon, and converts
the four-momentum given by~\eqref{eq:vector_n_general} into four
decoupled ordinary differential equations.
Hence, we first construct
\begin{align}
    \KNr&(\minoT,\Ktheta) \,, &
    \KNtheta&(\minoT,\Ktheta) \,,
    \label{eq:transformation_mino_to_r_theta}
\end{align}
and 
\begin{align}
    \CarterK\big(\KNr(\minoT,\Ktheta),\, \KNtheta(\minoT,\Ktheta)\big)
        = \KNa^2\sin^2\Ktheta \,,
\end{align}
and the adapted
coordinate $\Ktheta$ is also the initial angular value  on the horizon
($\Ktheta\horeq\KNtheta$). 

We have the Carter constant $\Cartereta<0$ and hence we follow the results 
for Vortical motion in \cite[Sec. III.B]{Gralla-2020}.
The angular potential $\KNS(\KNtheta)$ has four real roots
given by $\cos\KNtheta=\pm \sqrt{u_\pm}$, where 
for our choice of $\CarterK$ we have 
\begin{align}
    u_+ &= 1 \,,&
    u_- &= \frac{\KNa^2-\CarterK}{\KNa^2} = \cos^2\Ktheta \,.
\end{align}
Hence, the primitive function \cite[Eq.~(56)]{Gralla-2020}, i.e.
$\mathcal{G}_\KNtheta = \smallint \KKtheta^{\!\!\!-1} \d \KNtheta$,
can be simplified to 
\begin{align}
    \mathcal{G}_\KNtheta &= -\frac{\sign(\cos\KNtheta)}{\sqrt{\KNa^2\,u_-}}\,
    F\!\left(
    \arcsin \sqrt{\frac{\cos^2\KNtheta-u_{-}}{u_{+}-u_{-}}}
    \,\bigg|\, 1-\frac{u_+}{u_-}\right) ,
    \label{eq:privimite_Gtheta}
\end{align}
then
\begin{align}
    \begin{split}
    \cos\KNtheta_\obs &=
        h
        \sqrt{u_-}\, \dn\!\left(\sqrt{\KNa^2\, u_-}\, \big(\minoT
            +\mathcal{G}_\KNtheta^\source\big)\,\bigg|\,1-\frac{u_+}{u_-}\right) \\
                      &=
                      \cos\Ktheta \,\dn\!\Big(|\KNa\cos\Ktheta|
                          \left(\minoT+\mathcal{G}_\KNtheta^\source\right)
                      \,\Big|\,{-\tan^2\Ktheta}\Big) \,,
    \label{eq:cosThetaObs_vorti}
    \end{split}
\end{align}
where, as defined in \cite{Gralla-2020}, the sub/super-scripts $\obs$ and
$\source$ stand for observer or source, respectively, and we use the shortcut
$h = \sign(\cos\KNtheta)$, also used in~\cite{Gralla-2020}. 
In our case, \emph{the source} is a point on the outer black hole horizon defining
the new angular coordinate $\Ktheta$ (through $\KNtheta \horeq \Ktheta$)
, and \emph{the observer} is anywhere outside
(and therefore we will later omit the subscript $\obs$).

Later, we will see that the $\cal{G}_\KNtheta$ for our choice of parameters can
be simplified further, but at the moment we need only the fact that it vanishes
on the horizon:
\begin{align}
    \mathcal{G}_\KNtheta^\source &=\mathcal{G}_\KNtheta(\KNtheta=\Ktheta)=0 \,,
\end{align}

The structure of the roots of the radial potential $\KNR$ depends on the value
of $\CarterK$. For \remove{$\CarterK\in\langle 0,\CarterK_*\rangle$}\add{$\CarterK\in\langle 0,\CarterK_*)$},
i.e.\ from the axis to a certain angular coordinate where $\CarterK=\CarterK_*$,
the roots of the radial potential $\KNR$ consist of
two pairs of complex conjugate roots:
$\KNr_1$, $\KNr_2=\cconj{\KNr}_1$, 
$\KNr_3$, $\KNr_4=\cconj{\KNr}_3$. 
For $\CarterK=\CarterK_*$, the imaginary parts of $\KNr_1, \KNr_2$ vanish, i.e. $\KNr_1=\KNr_2=\KNr_*$, $\KNr_3=\KNr_*+\ii \KNr_i$, $\KNr_4=\cconj{\KNr}_3$,
and we have two real roots and a pair of
complex conjugate ones $\KNr_1<\KNr_2$, $\KNr_3$, $\KNr_4=\cconj{\KNr}_3$ for
\remove{$\CarterK\in\langle\CarterK_*, \KNa^2\rangle$}\add{$\CarterK\in(\CarterK_*, \KNa^2\rangle$}.
The reader is referred to \cite[Appendix B, Case (4, 3)]{Gralla-2020}
for a detailed discussion and definitions.

To solve for $\CarterK_*$, we need to find the root of the discriminant of $\KNR$
\begin{align}
\begin{split}
    0 &= \left(\massBH^2 - \KNa^2\right) \CarterK^3
    + \left(\KNa^4 + 30 \KNa^2 \massBH^2 - 27 \massBH^4\right) \CarterK^2 \\
      &\qquad{}- 96 \KNa^4 \massBH^2 \CarterK + 64 \KNa^6 \massBH^2 \,.
\end{split}
\end{align}
This equation has three real roots, and we consider the one for which
$\CarterK_*\in\langle 0, \KNa^2\rangle$.
Then, we can write down the radial antiderivative
$\mathcal{I}_\KNr=\smallint \KKr^{\!\!\!-1} \d \KNr$
given by \cite[Eq. (B97), (B101)]{Gralla-2020}
and \cite[Eq. (B67), (B71)]{Gralla-2020}
\begin{align}
    \mathcal{I}_\KNr &= 
    \begin{cases}
        \frac{2}{C+D}F\big(\arctan x_4(\KNr)+\arctan g_0\,|\,k_4\big) \,, \\[1ex]
        \hfill\text{for}\, \CarterK\in\langle 0,\CarterK_*\rangle \\[2ex]
        \frac{1}{\sqrt{AB}}\,F\big(\arccos x_3(\KNr)\,|\,k_3\big) \,, \\[1ex]
        \hfill\text{for}\, \CarterK\in\langle \CarterK_*, \KNa^2\rangle
    \end{cases}
    \label{eq:integral_Ir}
\end{align}
where the functions $A, B, C, D$ are defined in
\cite[Eq. (B57)]{Gralla-2020} and \cite[Eq. (B85)]{Gralla-2020},
\remove{and $k_3$ and $k_4$ are given by}
\add{and $x_3$, $k_3$, $x_4$, $k_4$ and $g_0$ are given by}
\cite[
\remove{Eq. (B59), (B87)}
\add{Eqs. (B58), (B59), (B83), (B87), (B88)}
]{Gralla-2020}.

Since we also have the expression for Mino time as
\begin{align}
    \minoT &= \mathcal{I}_\KNr^\obs-\mathcal{I}_\KNr^\source\,,
\end{align}
we can substitute this into Eq.~(\ref{eq:cosThetaObs_vorti})
and obtain explicit dependence $\KNtheta=\KNtheta(\Ktheta,\KNr)$ as
\begin{align}
    \cos\KNtheta &= \cos\Ktheta\, \dn\!\Big(|\KNa\cos\Ktheta|\,
        \left(\mathcal{I}_\KNr^\obs-\mathcal{I}_\KNr^\source\right)
    \,\Big|\, {-\tan^2\Ktheta}\Big) \,.
    \label{eq:cosThetaR_vorti}
\end{align}
Due to the analytic properties of $\dn$, this relation holds in
both hemispheres.
While explicitly determining $\KNtheta$, this is still an implicit equation%
~\eqref{eq:Carter_implicit1} for $\Ktheta$. Yet, it is useful when the expansion of
$\CarterK(\KNr,\KNtheta)=\KNa^2 \sin^2\Ktheta$ is constructed,
see Eq.~\eqref{eq:K_series_in_a_gen}, because it allows us to avoid solving
the differential equations for the coefficients that arise when
\eqref{eq:Carter_geodesic_cond} is solved using expansions directly.

Since $\NPn^\KNr$, defined in~\eqref{eq:vector_n_general_r}, is strictly negative
along the null congruence under consideration, $\KNr(\Kr)$ is monotone, and we
may invert the relation~\eqref{eq:vector_n_general_r} and write
\begin{align}
    \d \Kr &= \frac{\KNSigma}{\sqrt{\KNR}}\,\d \KNr \,.
\end{align}
Integrating this, we obtain
\begin{align}
    \Kr &= \int \frac{\KNr^2}{\sqrt{\KNR}}\,\d \KNr 
        + \int\frac{\KNa^2\cos^2\KNtheta(\KNr)}{\sqrt{\KNR}}\, \d \KNr \,.
\end{align}
The first integral leads to elliptic integrals,
see Appendix~\ref{sec:integrals} (where it is denoted as $\mathcal{J}_\KNr$),
whereas the second one can, taking into account
Eq.~\eqref{eq:cosThetaR_vorti}, be easily calculated by a change of variables
\begin{align}
    \frac{1}{\sqrt{\KNR}}\,\d \KNr &= \d \mathcal{I}_\KNr \,,
\end{align} 
and utilizing
\begin{align}
    \int \dn^2\!\big(y\,\mathcal{I}_\KNr\,\big|\,m\big)\, \d \mathcal{I}_\KNr &= 
    \frac{1}{y}\,\mathcal{E}\big(y\,\mathcal{I}_\KNr\,\big|\,m\big) \,,
\end{align}
where $y$ is an arbitrary constant.

This way, we explicitly get $\Kr(\KNr,\Ktheta)$ as
\begin{align}
    \Kr &= \mathcal{J}(\KNr)-\mathcal{J}(\KNrp) \nonumber\\
    &\quad+ |\KNa\cos\Ktheta|\; \mathcal{E}\Big(|\KNa\cos\Ktheta|\,
    \big(\mathcal{I}_\KNr(\KNr)-\mathcal{I}_\KNr(\KNrp)\big)
    \,\Big|\,-\tan^2\Ktheta\Big) \,.
    \label{eq:s_using_jacobi}
\end{align}

Altogether, \eqref{eq:cosThetaR_vorti} and \eqref{eq:s_using_jacobi}
provide an analytic form of the transformations
\begin{align}
    \Kr &= \Kr(\KNr,\Ktheta)\,,\\
    \KNtheta &= \KNtheta(\KNr,\Ktheta) \,.
\end{align}
Alas, their inversions seem to be impossible to obtain analytically.
The derivation of the first two functions already involves substantial
analytical complexity. Extending the same approach to $\tilde{\Kphi}$ would still
require the inversion of $\KNtheta$ and would not yield
additional closed-form insight. We therefore defer this function to
other approaches that follow.

\subsection{Numerical solution}\label{sec:numerical_solution}
As we mentioned, Eq.~\eqref{eq:Carter_implicit1}
is an implicit equation providing $\Ktheta$ for given $\KNr$, $\KNtheta$.
Even though both integrals in~\eqref{eq:Carter_implicit1} can be expressed
using elliptic integrals, in numerical evaluation, we do this only for 
$\mathcal{G}_\KNtheta$ given by \eqref{eq:privimite_Gtheta}.
With $\ru=1/\KNr$, the radial integral in~\eqref{eq:Carter_implicit1} can be written as
\begin{equation}
    \int\limits_{\KNrp}^\KNr
    \frac{\d \KNr'}{\KKr'}
    =
    \int\limits_\ru^{\ru_\text{p}}
    \frac{\d \ru}{\sqrt{\left(\KNa^2 \ru^2+1\right)^2-b^2 \ru^2
        \left(\KNa^2 \ru^2-2 \massBH \ru+1\right)}}\,,
\end{equation}
where $b=\KNa \sin\Ktheta$, and then evaluated efficiently using the Gauss--Legendre
quadrature. This way, we do not need to evaluate the roots of the radial potential  
\eqref{eq:radialPolynomialR} and avoid any problems with reduced precision of elliptic integrals \eqref{eq:integral_Ir} for certain moduli $k_3, k_4$.
Then, we equate the squares of both integrals appearing
in~\eqref{eq:Carter_implicit1}. This speeds up the root finding because
$\mathcal{G}_\KNtheta$ behaves as $\sim {|\Ktheta-\KNtheta|}^{\frac{1}{2}}$.

The coordinate transformations are based on solving ordinary differential
equations. Let us discuss here the more complicated situation when
the coordinates of~\eqref{eq:Kerr_null_coordinates} are given. Then we
first determine $\Ktheta$ and thus $\CarterK$. Subsequently, we solve the
Carter equations rewritten with $\ru$ as the independent variable. We use
quantities $z:=(\KNr-\KNrp)-\Kr$ and $w := |\sin(\Ktheta-\KNtheta)|^{1/2}$ to
allow $\KNr\to \infty$ and regularize the differential equation for $\KNtheta$
at the horizon. If we now use shortcuts $\KNR(\ru), \KNSigma(\ru,w), \KNtheta(w)$,
the set of differential equations reads
\begin{align}
    \frac{\d z}{\d \ru}&= \frac{\KNSigma-\sqrt{\KNR}}{\ru^2\sqrt{\KNR}} \,,
    \label{eq:CarterODEz}
    \\
    \frac{\d w}{\d \ru}&= - \KNa\, \frac{\cos\!\left(\Ktheta-\KNtheta\right)}
    {2\ru^2 \sqrt{\KNR}}\,
    {\big|\sin\!\left(\Ktheta+\KNtheta\right)\big|^{1/2}} \,,
    \\
    \frac{\d \tilde{\Kphi}}{\d \ru}&= \frac{\KNa}{\ru^2 \sqrt{\KNR}} \left( 1-\frac{\ru^2 \CarterK}{1+\KNa^2 \ru^2 + \ru^2\sqrt{\KNR}}\right) \,.
    \label{eq:CarterODEw}
\end{align}
Note that $\ru^2\sqrt{\KNR}$ as well as $\KNSigma-\sqrt{\KNR}$ behave as $\bigo(1)$
at infinity, i.e. $\ru\to 0$.
This form of the ordinary differential equation allows for efficient integration 
because 
\begin{align}
    \begin{split}
    &\cos^2\!\left(\Ktheta-\KNtheta\right)\big|\!\sin(\Ktheta+\KNtheta)\big| =\\
    &\quad{}\,\left(1-w^4\right)\!\left(\!\sqrt{1-w^4}\, \big|\!\sin(2\Ktheta)\big|
    -w^2 \cos (2 \Ktheta)\!\right) ,
    \end{split}
\end{align}
and 
\begin{align}
    \big|\!\cos\KNtheta\big| = \sqrt{1-w^4}\, \big|\! \cos \Ktheta \big|
    + w^2 \sin \Ktheta \,,
\end{align}
and therefore the
trigonometric functions must be evaluated only at the beginning of the integration.
With six steps of ninth-order Runge--Kutta method we get error
$\lesssim 10^{-15}$ for $|\KNa|<0.8$, $\tfrac{\KNrp}{2} < \KNr <\infty$.
To reach $|\KNa|\lesssim 1$, we need ten steps or must accept the error $\lesssim 10^{-13}$.
This numerical solution is used as a reference to estimate the errors of the series
expansions we construct in the subsequent parts of this paper and allows us to plot
various quantities for a broader range of parameters than the expansion would allow.

\begin{figure}
    \centering
    \figinput{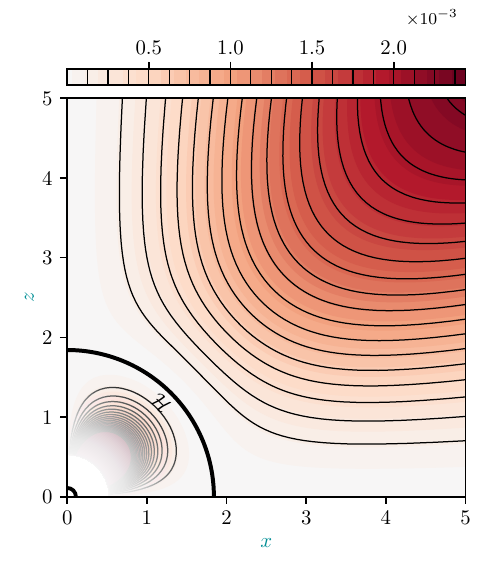}
    \caption{
        Contour plot adapted from~\cite[Fig.~3]{Kofron-2024} showing
        the difference $\CarterK - \KNa^2\sin^2\KNtheta$
        for values $\massBH = 1$ and $\KNa = 1/2$.
    }
    \label{fig:coor_function_K}
\end{figure}

\begin{figure}
    \centering
    \figinput{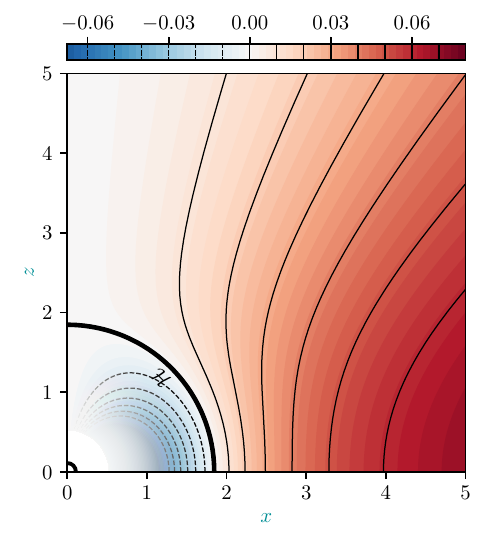}
    \caption{
        Contour plot demonstrating the difference of the Krishnan and Kerr null
        coordinates, $\left(\KNr - \KNrp\right) - \Kr$
        for values $\massBH = 1$ and $\KNa = 1/2$.
    }
    \label{fig:coor_function_s}
\end{figure}

\begin{figure}
    \centering
    \figinput{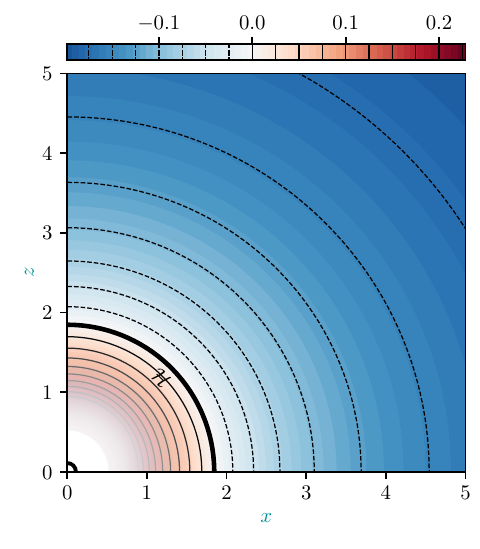}
    \caption{
        Contour plot demonstrating ($\KNr$, $\KNtheta$) dependence of
        the difference $\tilde{\Kphi} - \KNphi$
        for values $\massBH = 1$ and $\KNa = 1/2$.
    }
    \label{fig:coor_function_J}
\end{figure}

In Fig.~\ref{fig:coor_function_K}--\ref{fig:coor_function_J} the functions
$\CarterK$, $\Kr$ and $\tilde{\Kphi}$ governed by Eqs.%
~\eqref{eq:equation_for_s}--\eqref{eq:equation_for_J} are illustrated
using a numerical solution. To better highlight the relevant behavior,
the functions are shown through derived expressions.


\subsection{Series expansion around the horizon}
While there are obvious advantages to an analytical solution from the
previous section, the Jacobi elliptic functions, it is given by, can be
unwieldy. For the quasi-local isolated horizon, the close neighborhood
of the horizon itself is the section of the space-time that is of the most interest.
Therefore, we may construct a series expansion about the horizon of the form
\begin{equation}
    \CarterK = 
    \KNa^2\sin^2\KNtheta
    + \sum_{j = 1}^\infty\left(\KNr-\KNrp\right)^j\widetilde{C}^\CarterK_j(\KNtheta)
    \,.
\end{equation}
However, since $\CarterK\in\langle 0, \KNa^2\rangle$,
we will instead use a different series expansion, hoping for good convergence
at infinity. We may write
\begin{equation}
    \CarterK = \KNa^2\sin^2\KNtheta
    + \sum_{j = 1}^\infty\left(1-\frac{\KNrp}{\KNr}\right)^j\CCarterK_j(\KNtheta)
    \,.
\end{equation}

Solving Eq.~\eqref{eq:equation_for_K} order by order, we get 
\remove{(omitting the obvious dependence on $\KNtheta$):}
\add{(writing $\CCarterK_j$ instead of $\CCarterK_j(\KNtheta)$ for simplicity):}
\begin{subequations}
\begin{align}
    \CCarterK_1 &= 0 \,, \\
    \CCarterK_2 &= \frac{\KNa^4 \sin^2(2\KNtheta)}{16\massBH^2} \,, \\
    \CCarterK_3 &= \frac{\KNa^4 \sin^2(2\KNtheta)}{256\massBH^4}
    \,\KNrm\Big(16\massBH + 2\left(\KNrm - \KNrp\right) \sin^2\KNtheta\Big)
        \,.
\end{align}
\end{subequations}

An equation analogous to~\eqref{eq:equation_for_K} is valid for
$\Ktheta$ which is connected to $\CarterK$ by its choice on the horizon.


As in the case of $\CarterK$, we will seek a solution to Eq.%
~\eqref{eq:equation_for_s} in the form of a series expansion. In this case,
in the form
\begin{align}
    \Kr = \left(\KNr - \KNrp\right)
    + \sum_{j = 1}^\infty \left(1 - \frac{\KNrp}{\KNr}\right)^j \CKr_{j}(\KNtheta)
    \,,
\end{align}
where the term $\KNr - \KNrp$ could, of course, be absorbed into the sum,
but we keep it separate to indicate that at infinity $\Kr \sim \KNr$, as
expected. The coefficient $\CKr_0$ has been set to $0$ so that the affine
parameter is zero on the horizon.
\begin{subequations}
\begin{align}
    \CKr_1 &= - \frac{\KNa^2 \sin^2\KNtheta}{2\massBH} \,, \\
    \begin{split}
    \CKr_2 &= - \frac{\KNa^2 \sin^2\KNtheta}{32\massBH^3}
        \Big(4\KNa^2 + 5\KNrmSQ - \KNrp\!^2 \\
                     &\qquad\qquad\qquad\qquad{}- \KNrm\left(\KNrm - \KNrp\right) \sin^2\KNtheta\Big)
        \,.
    \end{split}
\end{align}
\end{subequations}


Yet again, we suppose a series expansion (in coordinates $\KNr$ and $\KNtheta$)
in the form
\begin{equation}
    \tilde{\Kphi} = \KNphi + \sum_{j = 0}^\infty\left(1-\frac{\KNrp}{\KNr}\right)^j\CJ_j(\KNtheta)
    \,.
\end{equation}
and solve Eq.~\eqref{eq:equation_for_J}.
Moreover, by Eq.~\eqref{eq:Jphi_hor_cond}, we have that
$\CJ_0 = 0$. The higher order coefficients are
\begin{subequations}
\begin{align}
    \CJ_1 &= -\frac{\KNa}{16\massBH^2}
    \Big(3 \KNrm + 4\KNrp + \KNrm \cos(2\KNtheta)\Big)
    \,, \\
    \begin{split}
    \CJ_2 &= -\frac{\KNa}{2\,048\massBH^4}
        \KNrm\Big(89\KNrmSQ + 247\KNrp\KNrm + 176\KNrp\!^2 \\
                    &\qquad{}+12\left(3\KNrmSQ + \KNrp\KNrm - 4\KNrp\!^2\right)\cos(2\KNtheta) \\
                    &\qquad{}+3\KNrm\left(\KNrm-\KNrp\right)\cos(4\KNtheta)\Big)
        \,.
    \end{split}
\end{align}
\end{subequations}

\begin{figure}
    \centering
    \figinput{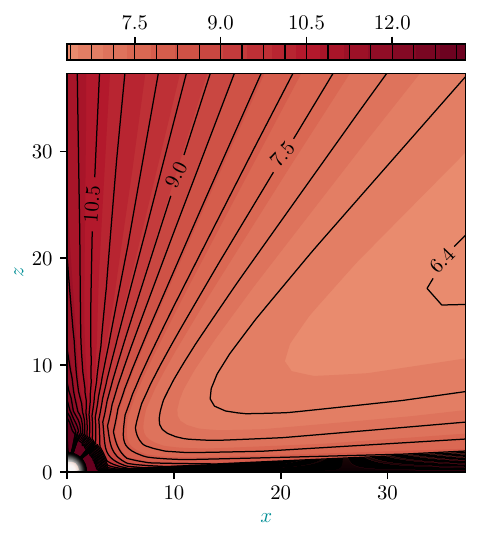}
    \caption{
        The number of digits correctly represented by the series expansion
        in the radial direction
        as estimated by~\eqref{eq:series_check_r}. The black hole is taken
        with mass $\massBH = 1$ and rotational parameter $\KNa = 0.5$.
        The series expansion used is up to the tenth order.
    }
    \label{fig:series_check_r}
\end{figure}

In the previous section, we showed how to construct a numerical solution. We can
use it to deduce the precision we achieve using the series expansion.
In Fig.~\ref{fig:series_check_r}, we compare the convergence of the solution through
the series expansion (up to the order $j$) and the numerical solution as given by
\begin{align}
    \Delta_\CarterK^j(\KNr,\KNtheta)\equiv -\log_{10}\left|
    \frac{\CarterK_j^{(\text{ser})}(\KNr,\KNtheta)-\CarterK^{(\text{num})}(\KNr,\KNtheta)}
    {\CarterK^{(\text{num})}(\KNr,\KNtheta)}
    \right| .
    \label{eq:series_check_r}
\end{align}
Hence, the value in the plot is the number of digits the series represents
correctly.
\begin{figure}
    \centering
    \figinput{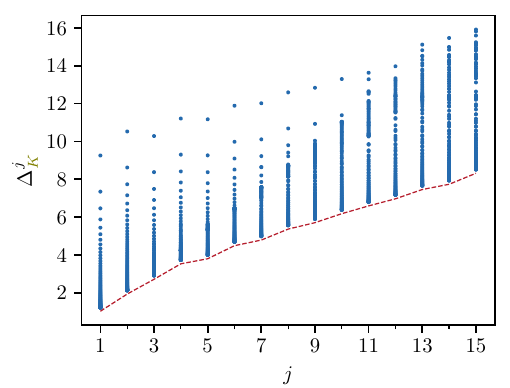}
    \caption{
        The function $\Delta_\CarterK^j$ as defined in Eq.~\eqref{eq:series_check_r} 
        for radial series expansion at $20\KNrp$ with $\massBH=1$, $\KNa=1/2$ evaluated
        for $100$ different values (columns of blue dots)
        of $\KNtheta \in (0, \Cpi/2)$ (at Chebyshev nodes) for each
        order $j$ on the horizontal axis.  The number on
        vertical axis corresponds to the number of digits correctly represented by the
        series expansion and the overall precision is given by the lower edge
        denoted by red dashed line.
    }
    \label{fig:series_check_r_orders}
\end{figure}
Insight into the convergence behavior of individual terms in the
series can be found in Fig.~\ref{fig:series_check_r_orders}.

\subsection{Slow rotation series expansion}
Let us again start with the Carter constant and a series expansion in the
rotational parameter $\KNa$:
\begin{equation}
    \CarterK = 
    \sum_{j = 0}^\infty \KNa^{2j} \caCarterK_{2j}(\KNr, \KNtheta)
    \,,
    \label{eq:K_series_in_a_gen}
\end{equation}
where we omitted odd powers since $\CarterK$ appears as the leading term under
a square root, and the zeroth-order term is excluded to enforce reflective
symmetry across the equatorial plane, \cite{Scholtz-2017}.

Solving Eq.~\eqref{eq:equation_for_K} we get the
\add{first non-zero} coefficients 
\remove{(again omitting the obvious dependence on $\KNr$ and $\KNtheta$)}
\add{(again writing $\caCarterK_j$ instead of $\caCarterK_j(\KNr,\KNtheta)$ for brevity)}
\begin{subequations}
\begin{align}
    \eqc\caCarterK_2 &\eqc=  \sin^2\KNtheta \,, \\
    \caCarterK_4 &=  \frac{\left(\KNr-2\massBH\right)^2 
        \sin^2(2\KNtheta)}{16\massBH^2\KNr^2} 
        \,, \\
    \begin{split}
    \caCarterK_6 &=  \frac{\left(r-2\massBH\right)
        \sin^2(2\KNtheta)}{512\massBH^4\KNr^5} 
        \Big(16\massBH^4 + 32\massBH^3\KNr + 11\KNr^4 \\
                                &\qquad{}- \left(\KNr-2\massBH\right)^2
        \left(4\massBH^2+12\massBH\KNr - 5\KNr^2\right) \cos(2\KNtheta)\Big)
        \,.
    \end{split}
\end{align}
\end{subequations}

Equation~\eqref{eq:equation_for_s} is solved using the series expansion
\begin{equation}
    \Kr = \sum_{j = 0}^\infty \KNa^{j} \caKr_j(\KNr, \KNtheta)
    \,.
\end{equation}
The first non-zero coefficients are
\begin{subequations}
\begin{align}
    \caKr_0 &=  \KNr - 2\massBH \,, \\
    \caKr_2 &=  
        \frac{5 + 3\cos(2\KNtheta)}{16\massBH} 
        + \frac{8\left(\KNr+\massBH\right)\sin^2\KNtheta}{16\KNr^2} \,.
\end{align}
\end{subequations}

Lastly, the coordinate $\tilde{\Kphi}$ is given by Eq.%
~\eqref{eq:equation_for_J}. The series expansion
\begin{equation}
    \tilde{\Kphi} = \sum_{j = 0}^\infty \KNa^{j} \caJ_j(\KNr, \KNtheta)
    \,,
\end{equation}
results in non-zero coefficients as
\begin{subequations}
\begin{align}
    \caJ_1 &=  \frac{1}{\KNr} - \frac{1}{2\massBH} \,, \\
    \caJ_3 &=  
        -\frac{29 + 3\cos(2\KNtheta)}{384\massBH^3}
        -\frac{3\massBH + 8\KNr - 3\cos(2\KNtheta)}{24\KNr^4} \,.
\end{align}
\end{subequations}

\begin{figure}
    \centering
    \figinput{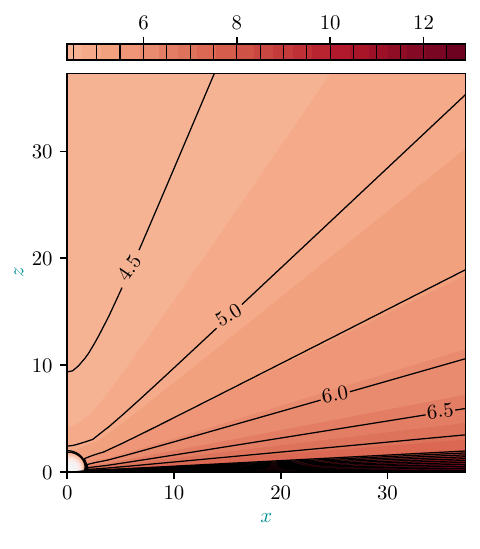}
    \caption{
        The number of digits correctly represented by the series expansion
        in the rotational parameter $\KNa$
        as estimated by~\eqref{eq:series_check_r}. The black hole is taken
        with mass $\massBH = 1$ and rotational parameter $\KNa = 0.5$.
        The series expansion used is up to the tenth order.
    }
    \label{fig:series_check_a}
\end{figure}

\begin{figure}
    \centering
    \figinput{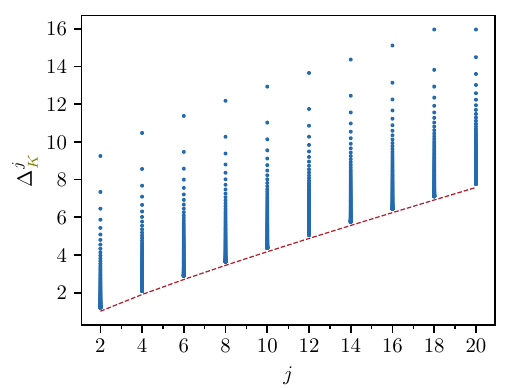}
    \caption{
        The function $\Delta_\CarterK^j$ as defined in Eq.~\eqref{eq:series_check_r} 
        for series expansion in $\KNa$
        at $20\KNrp$ with $\massBH=1$, $\KNa=1/2$ evaluated
        for $100$ different values (columns of blue dots)
        of $\KNtheta \in (0, \Cpi/2)$ (at Chebyshev nodes) for each
        order $j$ on the horizontal axis.  The number on
        vertical axis corresponds to number of digits correctly represented by the
        series expansion and the overall precision is given by the lower edge
        denoted by the red-dashed line.
    }
    \label{fig:series_check_a_orders}
\end{figure}

In complete analogy to Figs.~\ref{fig:series_check_r} and
\ref{fig:series_check_r_orders}, we show the achieved precision
of the series in $\KNa$ in Figs.~\ref{fig:series_check_a} and
\ref{fig:series_check_a_orders}.

In Appendix~\ref{sec:NP_scalars_a_series}, we give the
series expansions of the Newman--Penrose spin coefficients and
Weyl scalars in the Krishnan coordinates. Moreover, we illustrate 
both the most important scalars and the effect of black hole rotation
on those scalars.

Interestingly, using this approximation, it is actually possible to follow the
original approach of~\cite{Scholtz-2017}, now using the revised choice of $\CarterK$
\eqref{eq:K_is_a2_sin2_theta}, and to apply consecutive Lorentz transformations,
thereby circumventing the need for the affine parameter
$\Kr$ in the search for a geometrically correct tetrad. 
In fact, this was the approach taken initially, and its results
motivated the subsequent investigation.
We provide a brief recipe for such an approach in
Appendix~\ref{sec:slow_rotation_lorentz}.

\section{Conclusion}
    We analyzed the non-twisting geodesic congruence proposed in%
    ~\cite{Kofron-2024} for describing the isolated horizon of the
    Kerr black hole, employing the Newman--Penrose tetrad introduced in%
    ~\cite{Krishnan-2012}.
    The refined congruence is characterized by a Carter constant that
    remains invariant along each geodesic yet varies across the manifold.
    By applying the method for constructing parallel-propagated frames from%
    ~\cite{Kubiznak-2009}, we obtained the complete Newman--Penrose tetrad
    throughout the entire Kerr space-time.

    The resulting tetrad depends on several quantities given
    by implicit differential equations. These include the Carter constant itself,
    (minus) the affine parameter of the non-twisting geodesic congruence, and
    coordinate transformation parameters. 
    Since these objects encode the core computational complexity of the problem,
    we present multiple strategies for their evaluation, each with distinct
    benefits and limitations.
    We provide a fully analytical approach based on an extensive study of null
    geodesics in Kerr space-time \cite{Gralla-2020}, although its reliance on
    Jacobi elliptic functions often makes it rather unwieldy, and the final
    coordinate transformations cannot be inverted. As an alternative, we found
    radial series expansions that yield highly accurate approximations in the
    immediate vicinity of the horizon, as well as series expansions in the
    rotational parameter, which remain valid across the entire domain provided
    the black hole is sufficiently non-extremal. Finally, we tackled the
    problem through a numerical integration scheme.

    Both series-based approaches preserve all isolated-horizon conditions at
    every order, ensuring that truncated solutions continue to represent
    legitimate near-equilibrium black holes. These truncated configurations
    can be viewed as controlled perturbations of the exact Kerr geometry.

    Overall, the present work yields an explicit and physically well-adapted
    construction that improves upon the formulations of%
    ~\cite{Scholtz-2017, Kofron-2024}. 

\begin{acknowledgments}
    David Kofro\v{n} acknowledges support from Grant GACR 23-07457S of the Czech
    Science Foundation.
    The authors recognize the use of \emph{xAct} package for
    \emph{Wolfram Mathematica} \cite{xAct}.
\end{acknowledgments}

\appendix

\section{Lorentz transformations in Newman--Penrose formalism}
    \label{app:np:lorentz_transformation}
    The Newman--Penrose version of the six-parameter Lorentz group is, in the
    Newman--Penrose formalism, given by two real valued and two complex valued
    one-parameter subgroups:
    \begin{itemize}
        \item Boosts (with real parameter $A$):
        \begin{align}
            \NPl^a &\mapsto A^2\, \NPl^a\,, &
            \NPn^a &\mapsto A^{-2}\, \NPn^a\,, &
            \NPm^a &\mapsto \NPm^a \,.
            \label{eq:NP_boost}
        \end{align}
        \item Spins (with real parameter $\chi$):
        \begin{align}
            \NPl^a &\mapsto \NPl^a \,, &
            \NPn^a &\mapsto \NPn^a \,, &
            \NPm^a &\mapsto \eu^{2\ii \chi} \NPm^a \,.
            \label{eq:NP_spin}
        \end{align}
        \item Null rotations about $\NPl^a$ (with complex parameter~$c$):
        \begin{align}
            \begin{aligned}
                \NPl^a &\mapsto \NPl^a \,, \\
                \NPn^a &\mapsto \NPn^a + c\, \NPm^a + \cconj{c}\, \cconj{\NPm}^a + \abs{c}^2 \NPl^a \,, \\
                \NPm^a &\mapsto \NPm^a + \cconj{c}\, \NPl^a \,.
            \end{aligned}
            \label{eq:NP_rotation_l}
        \end{align}
        \item Null rotations about $\NPn^a$ (with complex parameter~$d$):
        \begin{align}
            \begin{aligned}
                \NPl^a &\mapsto \NPl^a + \cconj{d}\, \NPm^a + d\, \cconj{\NPm}^a + \abs{d}^2 \NPn^a \,, \\
                \NPn^a &\mapsto \NPn^a \,, \\
                \NPm^a &\mapsto \NPm^a + d\, \NPn^a \,.
            \end{aligned}
            \label{eq:NP_rotation_n}
        \end{align}
    \end{itemize}

\section{Integrals}\label{sec:integrals}
In Sec.~\ref{sec:coordinates}, we needed to calculate the following integrals
\begin{align}
    \mathcal{I}_r&= \int \frac{1}     {\sqrt{\KNR}} \,\d \KNr  \,,&
    \mathcal{J}_r&= \int \frac{\KNr^2}{\sqrt{\KNR}} \,\d \KNr  \,,
\end{align}
containing the square root of $4^\text{th}$ order polynomial
\begin{equation}
    \KNR=\left(\KNr-\KNr_1\right)\left(\KNr-\KNr_2\right)
    \left(\KNr-\KNr_3\right)\left(\KNr-\KNr_4\right) \,.
\end{equation}
The roots moreover satisfy
\begin{align}
    \KNr_1+\KNr_2+\KNr_3+\KNr_4 &=0 \,,
\end{align}
see \cite[Eq.~(96)]{Gralla-2020}.

\newcommand{\elphi}{\phi}
\newcommand{\nm}{n_\text{m}}
\newcommand{\np}{n_\text{p}}
For the possible values of the roots we find the following result
\begin{widetext}
\begin{align}
    \mathcal{I}_r &= \frac{\mathcal{P}}{\sqrt{\KNR}}\frac{2\KNr_{41}}{\KNr_{13}\,\KNr_{34}}
    \,F(\phi_\KNr,m_\KNr) \,,\\[1ex]
    \mathcal{J}_r &= \frac{\mathcal{P}}{\sqrt{\KNR}} \left[\frac{\KNr_{42}}{\KNr_{34}}
        \,E(\phi_\KNr,m_\KNr)+\frac{\KNr_1\KNr_{34}+\KNr_3(\KNr_3+\KNr_4)}{\KNr_{13}\,\KNr_{34}}
        \,F(\phi_\KNr,m_\KNr)\right]+\frac{\sqrt{\KNR}}{\KNr-\KNr_3} \,,
\end{align}
defining
\begin{align}
    \mathcal{P} &= \left(\KNr-\KNr_3\right)^2
    \KNr_{41}
    \,\sqrt{\frac{\left(\KNr-\KNr_4\right)\KNr_{13}}{\left(\KNr-\KNr_3\right)\KNr_{14}}}
    \,\sqrt{\frac{\left(\KNr-\KNr_1\right)\KNr_{34}}{\left(\KNr-\KNr_3\right)\KNr_{14}}}
    \,\sqrt{\frac{\left(\KNr-\KNr_2\right)\KNr_{34}}{\left(\KNr-\KNr_3\right)\KNr_{24}}} \label{eq:calP}\,,
\end{align}
where
\begin{align}
    \KNr_{ij} &= \KNr_i - \KNr_j \,, \\[1ex]
    m_\KNr &=\frac{\KNr_{32}\,\KNr_{41}}{\KNr_{31}\,\KNr_{42}}\,, \\[1ex]
    \phi_\KNr &= \arcsin\sqrt{\frac{\left(\KNr-\KNr_4\right)\KNr_{13}}
    {\left(\KNr-\KNr_3\right)\KNr_{14}}} \,.
\end{align}
It may be tempting to simplify (cancel out some terms) in $\mathcal{P}$ defined in Eq.~(\ref{eq:calP}), but it must be kept in mind that the square roots are complex numbers (the principal branch is considered for our calculations).

Finally, we return to the solution of Eq~\eqref{eq:phi_eqs_equator}. It is
given as
\begin{align}
    \begin{split}
        \mathcal{I}_{\tilde{\Kphi}} &= 
        \KNa\, \frac{\ln\!\big(1-\frac{\KNr}{\KNrp}\big) - \ln\!\big(1-\frac{\KNr}{\KNrm}\big)}{\KNrm - \KNrp} +
        \frac{4\KNa\massBH\KNr \left(\KNr - \KNr_4\right) \KNr_2 \sqrt{
                \frac{\left(\KNr - \KNr_3\right) \KNr_2 }
                {\left(\KNr - \KNr_2\right) \KNr_3 }}}
            {\left(\KNrm - \KNrp\right) \sqrt{\KNR}\, \KNr_4\, \KNDelta(\KNr_2) \sqrt{
                    \frac{\KNr\left(\KNr - \KNr_4\right) \KNr_{42} \KNr_2 }
                {\left(\KNr - \KNr_2\right)^2 \KNr_4^2 }}}
        \Big[\!
            \left(\KNrp-\KNrm\right) F(\elphi\,|\,m)\\
            &\qquad\qquad\qquad{}
            -\left(\KNrp - \KNr_2\right) \Pi(n(\KNrm);\,\elphi\,|\,m) 
            +\left(\KNrm - \KNr_2\right) \Pi(n(\KNrp);\,\elphi\,|\,m)
        \Big] 
        \,,
    \end{split}
    \label{eq:phi_eqs_equator_sol}
\end{align}
\end{widetext}
where $\Pi(n\,;\,\elphi\,|\,m)$ is the incomplete elliptic integral of the third kind
and $\KNr_i$ are the four roots of
\begin{equation}
    \KNR = \KNr\left(\KNr^3 + \KNr \KNrm \KNrp + \KNrm \KNrp \left(\KNrm - \KNrp\right)\right) = 0 \,,
\end{equation}
which have the structure
\begin{align}
    \KNr_1 &= 0 \,, &
    \KNr_2 &= -x - \cconj{x} \,, &
    \KNr_3 &= x \,, &
    \KNr_4 &= \cconj{x} \,,
\end{align}
where $\Re{x} > 0$ and $\Im{x} > 0$.
We also defined the following functions:
\begin{align}
    n(\KNr) &=\frac{\left(\KNr-\KNr_2\right) \KNr_{41}}{\left(\KNr-\KNr_1\right) \KNr_{42}}\,, \\
    \elphi &= \arcsin\sqrt{\frac{1}{n(\KNr)}} \,, \\
    m &= \frac{\KNr_{32}\, \KNr_{41}}{\KNr_{31}\, \KNr_{42}} \,.
\end{align}

This primitive function $\mathcal{I}_{\tilde{\Kphi}}$ is a smooth real valued
function on $\KNr\in\langle 0,\infty)$ (note, that there is no absolute value
in the logarithms; but there are elliptic integrals with complex parameters
as well). The manifest logarithmic singularities at $\KNrm$, $\KNrp$ are
rectified by the logarithmic singularities of elliptic integrals
$\Pi(n(\KNrm);\phi\,|\,m)$, $\Pi(n(\KNrp);\phi\,|\,m)$, which have branch
points on the horizons (since for fixed $n$, $m$ the function
$\Pi(n;\elphi\,|\,m)$ has a branch point for $z=\pm\arcsin(1/\sqrt{n})$).
In order to obtain a regular part (and thus the
definite integral
$\mathcal{I}_{\tilde{\Kphi}}(\KNr)-\mathcal{I}_{\tilde{\Kphi}}(\KNrp)$
which enters the coordinate transformation
$\KNphi=\tilde{\Kphi} + \mathcal{I}_{\tilde{\Kphi}}|_\KNrp^\KNr$,
one has to write the elliptic integrals $\Pi$ in terms of symmetric Carlson
integrals \cite{NIST:DLMF} and use  the results of \cite{Carlson-1993},
in particular \cite[Eq. (43)]{Carlson-1993}.
Since this computation is demanding (and giving us only $\tilde{\Kphi}$
in equatorial plane) and we obtain $\tilde{\Kphi}$ by other
methods (everywhere), we do not elaborate further.

As well, we understand that we could have used the fact that $\KNr_1=0$ in the
formulas, but they exhibit some form of symmetry if the $\KNr_1$ is kept
unevaluated.


\section{Tetrad transformations}
\label{app:metric_functions}
The final tetrad~\eqref{eq:general_tetrad},
discussed in Sec.~\ref{sec:non-twisting_tetrad}, obtained
from the parallel-propagated tetrad~\eqref{eq:PP_tetrad} is explicitly given by
the following metric functions:
\begin{subequations}
\begingroup
\allowdisplaybreaks
\begin{align}
    \begin{split}
        \TU &= \TU_\PP - \frac{\KNrp\!^2 + \KNa^2 - \CarterK}{2\CarterK} \\
            &\qquad{}- \frac{\ii\,\KNrp - \ethetaD \sqrt{\KNa^2 - \CarterK}}{\sqrt{2\CarterK}}\, \TOmega_\PP \\
            &\qquad{}+ \frac{\ii\,\KNrp + \ethetaD \sqrt{\KNa^2 - \CarterK}}{\sqrt{2\CarterK}}\, \cconj{\TOmega}_\PP
        \,, 
    \end{split} \\
    \begin{split}
    \TX^2 &= \TX^2_\PP 
            - \frac{\ii\,\KNrp - \ethetaD \sqrt{\KNa^2 - \CarterK}}{\sqrt{2\CarterK}}\, \Txi^2_\PP \\
          &\qquad{}+ \frac{\ii\,\KNrp + \ethetaD \sqrt{\KNa^2 - \CarterK}}{\sqrt{2\CarterK}}\, \cconj{\Txi}^2_\PP
    \,,
    \end{split} \\
    \begin{split}
        \TX^3 &= \TX^3_\PP - \frac{\KNa}{\KNrp\!^2 + \KNa^2} \\
              &\qquad{}- \frac{\ii\,\KNrp - \ethetaD \sqrt{\KNa^2 - \CarterK}}{\sqrt{2\CarterK}}\, \Txi^3_\PP \\
              &\qquad{}+ \frac{\ii\,\KNrp + \ethetaD \sqrt{\KNa^2 - \CarterK}}{\sqrt{2\CarterK}}\, \cconj{\Txi}^3_\PP
          \,,
    \end{split} \\
    \TOmega &= \sqrt{\frac{\KNrp + \ii \KNa \cos\Ktheta}
    {\KNrp - \ii \KNa \cos\Ktheta}} \left(
        \TOmega_\PP + \frac{\ii\,\KNrp + \ethetaD \sqrt{\KNa^2 - \CarterK}}{\sqrt{2\CarterK}}
    \right) , \\
    \Txi^2 &= \sqrt{\frac{\KNrp + \ii \KNa \cos\Ktheta}
    {\KNrp - \ii \KNa \cos\Ktheta}}\,\Txi^2_\PP \,, \\
    \Txi^3 &= \sqrt{\frac{\KNrp + \ii \KNa \cos\Ktheta}
    {\KNrp - \ii \KNa \cos\Ktheta}}\,\Txi^3_\PP \,. 
\end{align}
\endgroup
\end{subequations}

\section{Slow rotation expansion approach}\label{sec:slow_rotation_lorentz}

In analogy with the approach of~\cite{Scholtz-2017},
we can construct a parallel-propagated tetrad by applying general Lorentz
transformations to the Kinnersley tetrad%
~\eqref{eq:Kinnersley_tetrad}
The transformation parameters, taken as a series expansion with coefficients
depending on $\KNr$ and $\KNtheta$, are determined by solving the (differential)
equations imposed by the conditions of parallel transport.

First, the vector $\NPn^a_\KIN$ is aligned with the chosen geodesic direction,
which can be, using the series expansion~\eqref{eq:K_series_in_a_gen}, written as
\begin{subequations}
\begin{align}
    \NPn^\KNv_\GEO &= - \frac{\sin^2\KNtheta}{2\KNr^2}\KNa^2 + \bigo(\KNa^3) \,, \\
    \NPn^\KNr_\GEO &= - 1 - \frac{\left(2\massBH + \KNr\right) \sin^2\KNtheta}{2\KNr^3}\KNa^2 + \bigo(\KNa^3) \,, \\
    \NPn^\KNtheta_\GEO &= - \frac{\left(2\massBH - \KNr\right) \sin(2\KNtheta)}{4\massBH\KNr^3} \KNa^2 + \bigo(\KNa^3) \,, \\
    \NPn^\KNphi_\GEO &= - \frac{1}{\KNr^2} \KNa + \bigo(\KNa^3) \,.
\end{align}
\label{eq:NPn_prototype}
\end{subequations}
To achieve this, both a boost~\eqref{eq:NP_boost}
with parameter $A$ and a null rotation about $\NPl^a$~\eqref{eq:NP_rotation_l}
parametrized by $c$ are required. These are analogies of%
~\eqref{eq:boost_gen} and \eqref{eq:rot_l_gen}.

Next, we need to ensure that the remaining three vectors are parallel propagated.
To this end, we can use the remaining two Lorentz transformations --- rotation
about $\NPn^a$ and spin, neither of which changes the already set vector $\NPn^a$.

While in Sec.~\ref{sec:non-twisting_tetrad} we first performed the rotation
about $\NPn^a$ and then the spin, it is more convenient to switch the order
in this case. The parallel propagation is connected to the spin coefficients
$\gamma$, $\Snu$, and $\Stau$. Since $\NPn^a$ is parallel propagated, $\Snu = 0$.
The spin enables us to set $\Sgamma = 0$ while leaving $\Snu$. 
Note that there is a natural freedom in the choice of the spin parameter, which
we can later use to ensure the Ashtekar choice of vectors $\NPm^a$ and $\cconj{\NPm}^a$.
While the following rotation about $\NPn^a$ changes $\Sgamma$, the change
vanishes for $\Snu = 0$, and we can set the parameter 
\remove{to zero $\Stau$}\add{such that $\Stau$ vanishes}.
However, we also need to inspect the spin coefficients $\Skappa$, $\Srho$ and
$\Ssigma$. In the case of these coefficients, we want to ensure they vanish
on the horizon.
The resulting tetrad is not only parallel propagated, but has the vectors
Lie transported on the horizon as it should.

The next step is to find the coordinate transformation to the adapted
coordinates. Yet again, we can express the new coordinates as a series expansion
with coefficients generally depending on all the old coordinates.
We also need a similar series for the old coordinates expressed in terms of the
new ones. We can find the coefficients as follows:
\begin{itemize}
    \item $\Kphi$ based on $\NPn^\Kphi = 0$ and $\NPl^\Kphi \horeq 0$,
    \item $\Ktheta$ based on $\NPn^\Ktheta = 0$ and $\NPl^\Ktheta \horeq 0$,
    \item $\Kv$ based on $\NPn^\Kv = 0$, $\NPm^\Kv = 0$ and $\NPl^\Kv \horeq 1$,
    \item $\Kr$ based on $\NPn^\Kr = -1$ and $\NPl^\Kr \horeq 0$.
\end{itemize}
Additional conditions are needed to fix the freedom in the new coordinates.
For the radial coordinate, we want to have
$\Kr \horeq 0$.
The polar coordinate is best set using relations from Sec.~\ref{sec:axial_IH},
namely \eqref{eq:axial_h_cond}. 

This leads to an adapted tetrad in adapted coordinates in the form of a series
expansion, and it is easy to compute everything else. As a closing note, we found
it more convenient to solve the equations for the new radial coordinate such that
$\tilde{\Kr} = \Kr + \KNrp$, and only then correct the expansions by the series
of $\KNrp$.

\begin{figure*}
    \centering
    \figinput{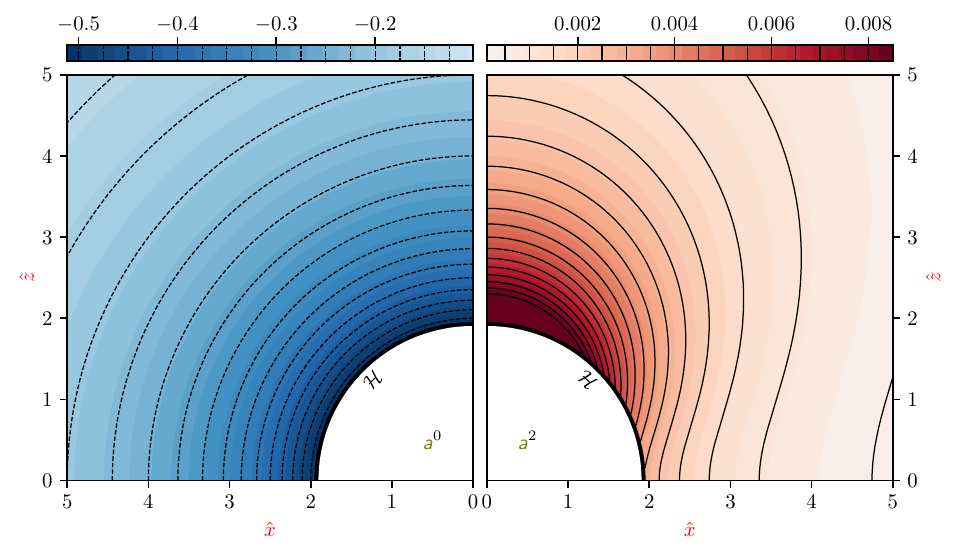} 
    \caption{
        The zeroth and second order terms of the spin coefficient $\Smu$.
        The order is indicated in the area ``under the horizon''.
        The mass of the black hole is set to $\massBH = 1$, and the rotational
        parameter is $\KNa = 0.3$. Note that both orders are displayed for the same
        quadrant and are mirrored solely for aesthetic purposes.
        This spin coefficient is the expansion of the congruence.
    }
    \label{fig:slow_Kerr_mu}
\end{figure*}
\begin{figure*}
    \centering
    \figinput{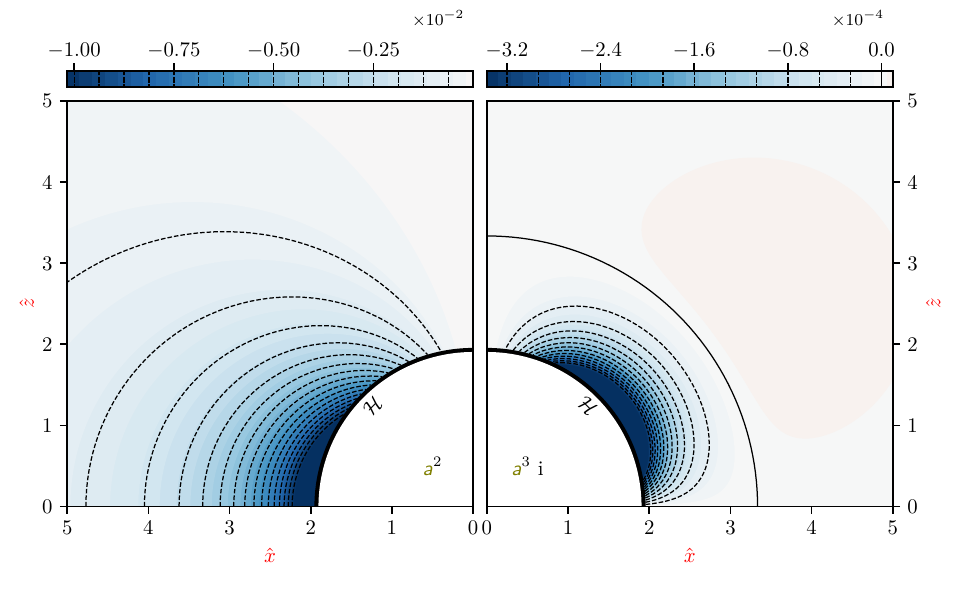} 
    \caption{
        The second and third order terms of the spin coefficient $\Slambda$.
        The third order is purely imaginary, which is indicated, together with the
        order, in the area ``under the horizon''.
        The mass of the black hole is set to $\massBH = 1$, and the rotational
        parameter is $\KNa = 0.3$. Note that both orders are displayed for the same
        quadrant and are mirrored solely for aesthetic purposes.
        This spin coefficient is the shear of the congruence.
    }
    \label{fig:slow_Kerr_lambda}
\end{figure*}
\begin{figure*}
    \centering
    \figinput{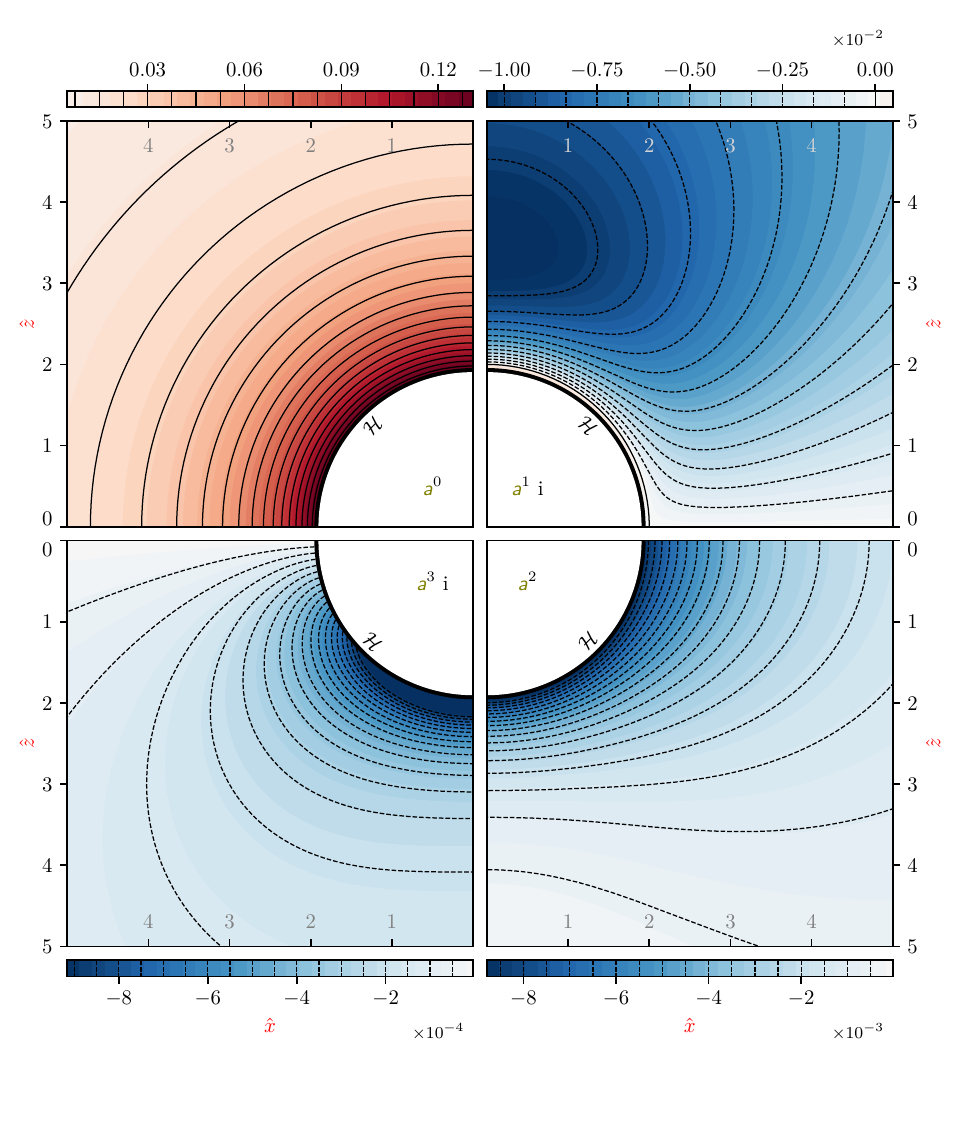} 
    \caption{The zeroth (top left) through third (clockwise) order terms
    of the spin coefficient $\Sepsilon$. The first and third orders are purely
    imaginary, which is indicated, together with the order, in the area
    ``under the horizon''.
    The mass of the black hole is set to $\massBH = 1$, and the rotational
    parameter is $\KNa = 0.3$. Note that all orders are displayed for the same
    quadrant and are mirrored solely for aesthetic purposes.}
    \label{fig:slow_Kerr_epsilon}
\end{figure*}
\begin{figure*}
    \centering
    \figinput{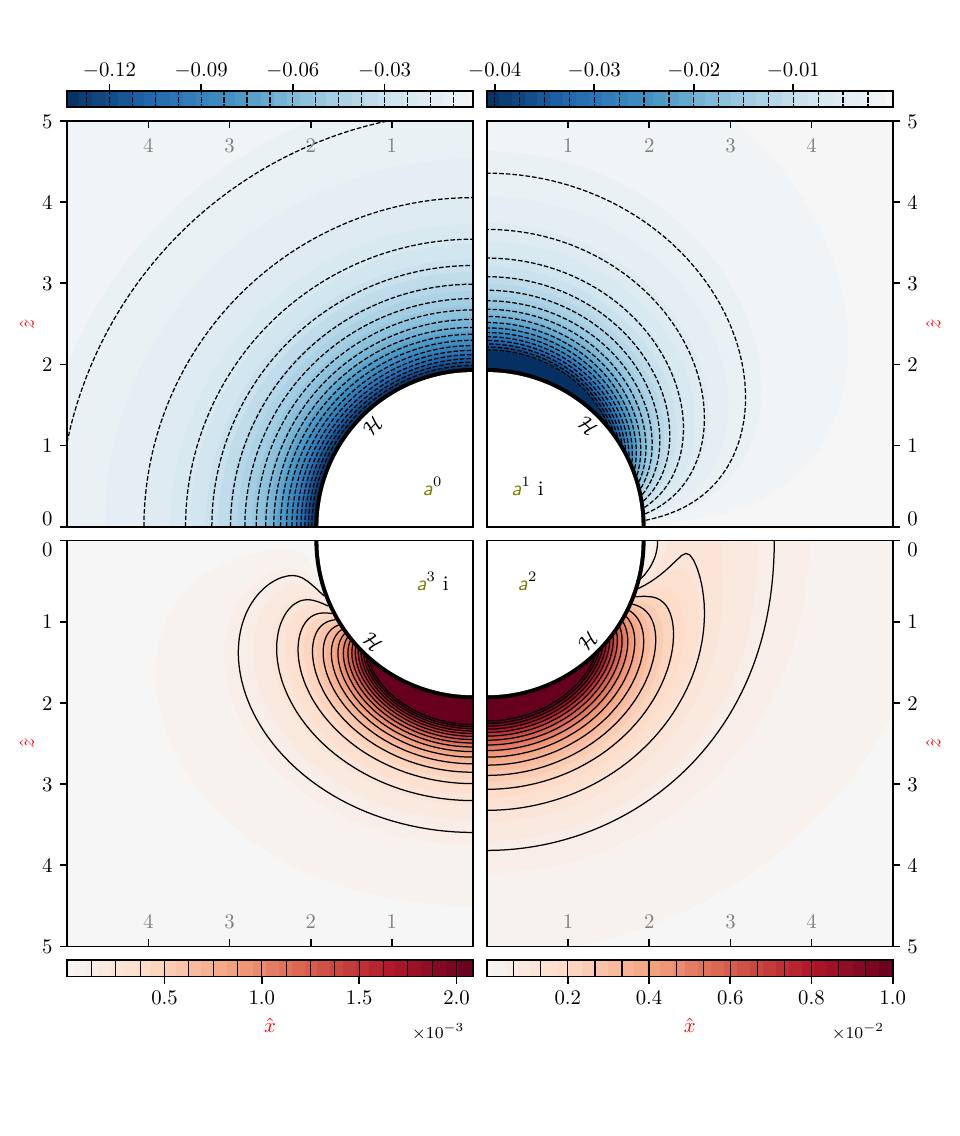} 
    \caption{The zeroth (top left) through third (clockwise) order terms
    of the Weyl scalar $\NPPsi_2$. The first and third orders are purely
    imaginary, which is indicated, together with the order, in the area
    ``under the horizon''.
    The mass of the black hole is set to $\massBH = 1$, and the rotational
    parameter is $\KNa = 0.3$. Note that all orders are displayed for the same
    quadrant and are mirrored solely for aesthetic purposes.}
    \label{fig:slow_Kerr_Psi2}
\end{figure*}
\begin{figure*}
    \centering
    \figinput{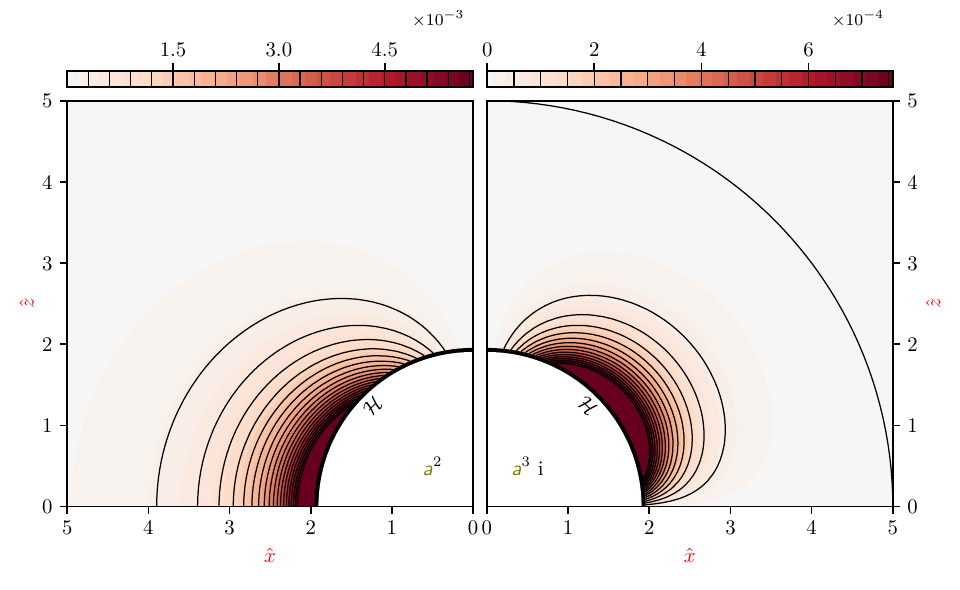} 
    \caption{The second and third order terms of the Weyl scalar $\NPPsi_4$.
    The third order is purely imaginary, which is indicated, together with the
    order, in the area ``under the horizon''.
    The mass of the black hole is set to $\massBH = 1$, and the rotational
    parameter is $\KNa = 0.3$. Note that both orders are displayed for the same
    quadrant and are mirrored solely for aesthetic purposes.}
    \label{fig:slow_Kerr_Psi4}
\end{figure*}
\vspace{1.5em}
\section[Newman--Penrose scalars series expansion in a]{Newman--Penrose scalars series expansion in $\boldsymbol{\KNa}$}\label{sec:NP_scalars_a_series}
    With the tetrad constructed, the Newman--Penrose scalars can be easily
    computed. The most relevant ones are presented in Figs.%
    ~\ref{fig:slow_Kerr_mu}--\ref{fig:slow_Kerr_Psi4}. Although the tetrad was
    primarily displayed up to second order due to its length, it was computed
    to third order and the same applies to the Newman--Penrose scalars and the
    figures. The figures are displayed in coordinates $\KNSEx$ and $\KNSEz$
    introduced by
    \begin{align}
        \KNSEx &= \left(\KNSr + \KNrp\right) \sin\KNStheta \,, &
        \KNSEz &= \left(\KNSr + \KNrp\right) \cos\KNStheta \,.
        \label{eq:euclid_kerr_ih_coor}
    \end{align}
    All figures are plotted for $\KNStheta \in (0, \Cpi/2)$ and while the
    even orders are identical in all quadrants due to symmetries, the odd
    orders have inverse values for $\KNSEz < 0$.

    Each successive order in $\KNa$ is significantly smaller than the
    preceding one. Instead of plotting the combined amplitude, which would obscure the
    contributions of higher-order terms, we show each order in $\KNa$ up to $\KNa^4$
    separately. Any order not shown in a given figure corresponds to
    a vanishing contribution for that scalar.
    
    The expressions for the spin coefficients are:
    \begin{widetext}
    \vspace{-0.5cm}
    \begingroup
    \allowdisplaybreaks
    \begin{align}
        \begin{split}
            \Skappa   &= - \frac{3\ii\Kr^2 \sin\Ktheta}
            {16\sqrt{2}\, \massBH \KrMM^3} \KNa 
                      + \frac{\Kr^2 \left(136\massBH^2 + 80\massBH\Kr
                      + 11 \Kr^2\right) \sin(2\Ktheta)}{128\sqrt{2}\,\massBH^2 \KrMM^5} \KNa^2 
                      + \bigo(\KNa^3)
            \,,
        \end{split}
        \\[1.5ex]%
        \begin{split}
            \Ssigma   &= \frac{\Kr^2 \left(40\massBH^2 + 24\massBH\Kr
            + 5 \Kr^2\right) \sin(2\Ktheta)}{128\massBH^2 \KrMM^5} \KNa^2 
            + \bigo(\KNa^3)
            \,,
        \end{split}
        \\[1.5ex]%
        \begin{split}
            \Srho     &= - \frac{\Kr}{2\KrMM^2} - \frac{3\ii
            \Kr\left(4\massBH + \Kr\right) \cos\Ktheta}{8\massBH\KrMM^3}\KNa \\
                      &\qquad{}+ \frac{
                \Kr^2 \left(6\massBH^2 + 2\massBH\Kr + \Kr^2\right)
                - 3\Kr \left(64\massBH^3 + 46\massBH^2 \Kr
            + 10 \massBH \Kr^2 + \Kr^3\right) \cos(2\Ktheta)
            }{64\massBH^2\KrMM^5} \KNa^2
            + \bigo(\KNa^3)
            \,, 
        \end{split}
        \\[1.5ex]%
        \begin{split}
            \Sbeta    &= \frac{\cot\Ktheta}{2\sqrt{2}\,\KrMM}
            - \frac{3\ii\sin\Ktheta}{8\sqrt{2}\,\massBH\KrMM}\KNa 
                      + \frac{2\KrMM^3 \cot\Ktheta
            + \left(8\massBH^3 + 16\massBH^2\Kr + 7\massBH\Kr^2 + \Kr^3\right)
            \sin(2\Ktheta)}{32\sqrt{2}\,\massBH^2\KrMM^4} \KNa^2
            + \bigo(\KNa^3)
            \,, 
        \end{split}
        \\[1.5ex]%
        \begin{split}
            \Salpha   &= -\frac{\cot\Ktheta}{2\sqrt{2}\,\KrMM}
            + \frac{3\ii\massBH\sin\Ktheta}{2\sqrt{2}\,\KrMM^3}\KNa 
                      - \frac{2\KrMM^3 \cot\Ktheta
            + \massBH \left(40\massBH^2 - 8\massBH\Kr + \Kr^2\right)
            \sin(2\Ktheta)}{32\sqrt{2}\,\massBH^2\KrMM^4} \KNa^2
            + \bigo(\KNa^3)
            \,,
        \end{split}
        \\[1.5ex]%
        \begin{split}
            \Sepsilon &= \frac{\massBH}{2\KrMM^2} - \frac{3\ii \Kr
            \left(4\massBH + \Kr\right) \cos\Ktheta}{16\massBH\KrMM^3}\KNa \\
                      &\qquad{}- \frac{256\massBH^4 + 352\massBH^3\Kr 
                      + 208\massBH^2\Kr^2 + 72\massBH\Kr^3 + 9\Kr^4 
            - 3\Kr \left(160\massBH^3 + 112\massBH^2 \Kr + 24\massBH\Kr^2 + 3\Kr^3\right)
            \cos(2\Ktheta)}{256\massBH^2\KrMM^5} \KNa^2
            + \bigo(\KNa^3)
            \,,
        \end{split}
        \\[1.5ex]%
        \begin{split}
            \Spi      &= \frac{3\ii \left(8\massBH^2 + 4\massBH\Kr + \Kr^2\right)
            \sin\Ktheta}{8\sqrt{2}\,\massBH\KrMM^3}\KNa
            - \frac{\left(32\massBH^3 - 24\massBH^2\Kr - 6\massBH\Kr^2 - \Kr^3\right) \sin(2\Ktheta)}
            {32\sqrt{2}\,\massBH^2\KrMM^4} \KNa^2
            + \bigo(\KNa^3)
            \,,
        \end{split}
        \\[1.5ex]%
        \begin{split}
            \Slambda  &= - \frac{\left(10\massBH^2 + 4\massBH\Kr + \Kr^2\right)
            \sin^2\Ktheta}{4\massBH\KrMM^4}\KNa^2
            + \bigo(\KNa^3)
            \,,
        \end{split}
        \\[1.5ex]%
        \begin{split}
            \Smu      &= - \frac{1}{\KrMM} - \frac{3\sin^2\Ktheta}{8\massBH\KrMM^2}\KNa^2 + \bigo(\KNa^3) 
            \,.
        \end{split}
    \end{align}
    \endgroup
    The remaining spin coefficients $\Stau$, $\Sgamma$ and $\Snu$ vanish.
    \end{widetext}

    The projections of the Weyl tensor are:
    \begingroup%
    \begin{subequations}%
    \begin{align}%
        \begin{split}%
            \NPPsi_0 &= \frac{3 \Kr^2 \left(8\massBH
            + 3\Kr\right)^2 \sin^2\Ktheta}{64 \massBH \KrMM^7} \KNa^2 + \bigo(\KNa^3)
            \,, 
        \end{split}%
            \\[1.5ex]%
        \begin{split}%
            \NPPsi_1 &= - \frac{3\ii\left(8\massBH + 3\Kr\right)
            \sin\Ktheta}{8\sqrt{2}\, \KrMM^5} \KNa \\
             &\qquad{}+ \frac{3\Kr \left(64\massBH^2 + 28\massBH\Kr
                + \Kr^2\right) \sin(2\Ktheta)}{32\sqrt{2}\,\massBH \KrMM^6} \KNa^2 \\
             &\qquad{}+ \bigo(\KNa^3)
            \,,
        \end{split}%
            \\[1.5ex]%
        \begin{split}%
            \NPPsi_2 &= -\frac{\massBH}{\KrMM^3} - \frac{3\ii \massBH\cos\Ktheta}{\KrMM^4}\KNa \\
                     &\qquad{}- \frac{3\left(\Kr^2 - \left(16\massBH^2 - 3\Kr^2\right)
                     \cos(2\Ktheta)\right)}{8\KrMM^6} \KNa^2 + \bigo(\KNa^3)
            \,,
        \end{split}%
            \\[1.5ex]%
        \begin{split}%
            \NPPsi_3 &= - \frac{3\ii\massBH\sin\Ktheta}{\sqrt{2}\,\KrMM^4} \KNa \\
                     &\qquad{}+ \frac{3\left(6\massBH - \Kr\right)\sin(2\Ktheta)}{4\sqrt{2}\,\KrMM^5} \KNa^2
            + \bigo(\KNa^3)
            \,,
        \end{split}%
            \\[1.5ex]%
        \begin{split}%
            \NPPsi_4 &= \frac{3\massBH\sin^2\Ktheta}{\KrMM^5}\KNa^2 + \bigo(\KNa^3)
            \,.
        \end{split}%
    \end{align}%
    \end{subequations}%
    \endgroup%

\clearpage

%

\end{document}